\documentclass[aps,pre,onecolumn,noshowpacs,superscriptaddress,groupedaddress,preprintnumbers,nofootinbib,showpacs,12pt]{revtex4}
\usepackage{epsfig,dsfont,amssymb,amsmath,amsthm,amsfonts,amsbsy,mathrsfs}
\usepackage{lipsum}
\usepackage{color}
\usepackage{graphicx}
\usepackage{amsmath}
\usepackage{amssymb}
\usepackage{subfigure}
\usepackage{float}
\usepackage{hyperref}
\usepackage{verbatim}
\usepackage{soul}
\setstcolor{red}

\renewcommand{\eqref}[1]{\textrm{Eq.}(\ref{#1})}

\newcommand {\be}{\begin{equation}}
\newcommand {\ee}{\end{equation}}

\begin{document}


\title{Free energy dissipation enhances spatial accuracy and robustness of Turing pattern in small reaction-diffusion systems}

\author{Dongliang Zhang}
\affiliation{The State Key Laboratory for Artificial Microstructures and Mesoscopic Physics, School of Physics, Peking University, Beijing 100871, China}
\author{Chenghao Zhang}
\affiliation{The State Key Laboratory for Artificial Microstructures and Mesoscopic Physics, School of Physics, Peking University, Beijing 100871, China}
\affiliation{Physics Department, University of Illinois, Urbana, IL61801}
\author{Qi Ouyang}
\affiliation{The State Key Laboratory for Artificial Microstructures and Mesoscopic Physics, School of Physics, Peking University, Beijing 100871, China}
\affiliation{Center for Quantitative Biology and Peking-Tsinghua Center for Life Sciences, AAIC, Peking University, Beijing 100871, China}
\author{Yuhai Tu}
\affiliation{IBM T. J. Watson Research Center, Yorktown Heights, New York 10598, USA}

\email{yuhai@us.ibm.com}

\begin{abstract}
Accurate and robust spatial orders are ubiquitous in living systems. In 1952, Alan Turing proposed an elegant mechanism for pattern formation based on spontaneous breaking of the spatial translational symmetry in the underlying reaction-diffusion system. Much is understood about dynamics and structure of Turing patterns. However, little is known about the energetic cost of Turing pattern. Here, we study nonequilibrium thermodynamics of a small spatially extended biochemical reaction-diffusion system by using analytical and numerical methods. We find that the onset of Turing pattern requires a minimum energy dissipation to drive the nonequilibrium chemical reactions. Above onset, only a small fraction of the total energy expenditure is used to overcome diffusion for maintaining the spatial pattern. We show that the positioning error decreases as energy dissipation increases following the same tradeoff relationship between timing error and energy cost in biochemical oscillatory systems. In a finite system, we find that a specific Turing pattern exists only within a finite range of total molecule number, and energy dissipation broadens the range, which enhances the robustness of the Turing pattern against molecule number fluctuations in living cells. These results are verified in a realistic model of the Muk system underlying DNA segregation in E. coli, and testable predictions are made for the dependence of the accuracy and robustness of the spatial pattern on the ATP/ADP ratio.  In general, the theoretical framework developed here can be applied to study nonequilibrium thermodynamics of spatially extended biochemical systems.
   
\end{abstract}

\maketitle
\clearpage

\section{Introduction}

Spatial order (regularity) and pattern formation are ubiquitous in 
living organisms. Examples can be found in all living organisms spanning a large range of spatial and temporal scales, which ranges from patterning in limb development~\cite{Raspopovic2014Digit} and feathers and hair in the skins of birds and mammals~\cite{Painter2012Towards} to phillotaxis in plants~\cite{Richard2006Phyllotaxis} to accurate positioning of the chromosomal origin of replication in bacteria~\cite{badr2015bacterial}. 
Pattern formation in systems far from equilibrium have been studied extensively in large physical systems such as fluid systems where the number of molecules is of the order of the Avogadro number (see Cross and Hohenberg~\cite{Cross1993Pattern} for a comprehensive review). 
However, living systems are governed by biochemical reactions with a relatively small number of molecules, thus the underlying dynamics is subject to large stochastic noise and fluctuations~\cite{Butler2011, Karig2018, Diego2018}. 
Yet, accuracy of the spatial pattern or structure is crucial for the proper function of the organism. This raises the important questions on how spatial accuracy is affected by the biochemical noise in living system, how living system controls the noise, and what is the energy cost for achieving higher spatial accuracy. These are the general questions we try to address in this paper in the context of Turing pattern in small systems.

Recently, there have been increasing interests in understanding the relationship between performance of biological functions and their energetic costs in various nonequilibrium biological systems such as ultrasensitive biological switch~\cite{Tu2008Switch}, sensory adaptation~\cite{Lan2012}, biochemical oscillation~\cite{Cao2015}, biochemical error correction~\cite{Sartori2015Thermodynamics}, gene regulation~\cite{Estrada2016Information}, and synchronization~\cite{Zhang2019Synch}. These studies applied the nonequilibrium thermodynamics approach~\cite{hill_1977, Qian2007, ge2010, Rao2016}, which was developed to treat spatially homogeneous systems where the spatial degrees of freedom are irrelevant or the underlying biochemical reactions are well stirred. 

In this paper, we aim to understand positional order and its thermodynamic cost in reaction-diffusion systems by first extending the nonequilibrium thermodynamics framework to spatially extended systems where transport of molecules and the associated energy cost are considered explicitly. 
We then use this extended theoretical framework to study nonequilibrium thermodynamics of a simplified reaction-diffusion model inspired by a realistic biological system where Turing pattern emerges as the system is driven away from equilibrium by increasing energy dissipation. In particular, we investigate how much energy is needed to generate and maintain the Turing pattern, how accuracy of the Turing pattern depend on the free energy dissipation, whether and how energy dissipation affects robustness of the Turing pattern against variations in key parameters such as the number of molecules in the system. Finally, we study a realistic biological system and propose experiments to test some of the predictions from our theoretical analysis. 

\section{Model and Analysis} 

\subsection{A simple biochemical reaction-diffusion model for Turing pattern}

To study thermodynamics of Turing pattern in biochemical systems, we used a modified 3-state reaction network model proposed by Murray and Sourjik~\cite{Murray2017} for studying DNA segregation. 
As shown in Fig.~1A, there are three species $X_1$, $X_2$, $X_3$ representing different forms (conformations) of the same protein complex. They can convert from one form to another in four reversible reactions with different transition rates as illustrated in Fig.~1B. In addition to three ``linear" reactions between all pairs of species, there is a ``nonlinear" auto-catalytic reaction where $X_1$ can convert to $X_2$  in the presence of two $X_2$:
\begin{equation}
X_1 \underset{k_{21}}{\stackrel{k_{12}}{\rightleftharpoons}} X_2,\;\; 
X_2 \underset{k_{32}}{\stackrel{k_{23}}{\rightleftharpoons}} X_3,\;\; X_3 \underset{k_{13}}{\stackrel{k_{31}}{\rightleftharpoons}} X_1,\;\; X_1+2X_2  \underset{\tilde{k}_{21}}{\stackrel{\tilde{k}_{12}}{\rightleftharpoons}} 3X_2,
\label{reaction}
\end{equation}
where $k_{ij}$ ($i\in[1,3]$, $j\in[1,3]$, $i\ne j$) are the reaction rates for the linear conversion reactions, and $\tilde{k}_{12(21)}$ are the rates for the reversible autocatalytic reaction. Note that although the topology of the reaction network is the same as in \cite{Murray2017}, a key difference is that all reactions in our model are reversible with non-zero forward and backward rates, which allows us to study thermodynamics of the system properly (see Supplementary Material (SM) for details of dynamical equations). The original 3-state model~\cite{Murray2017} considered the irreversible limit for the autocatalytic reaction ($\tilde{k}_{21}=0$).  

   \begin{figure}
   	\centering
\includegraphics[width=0.9\linewidth]{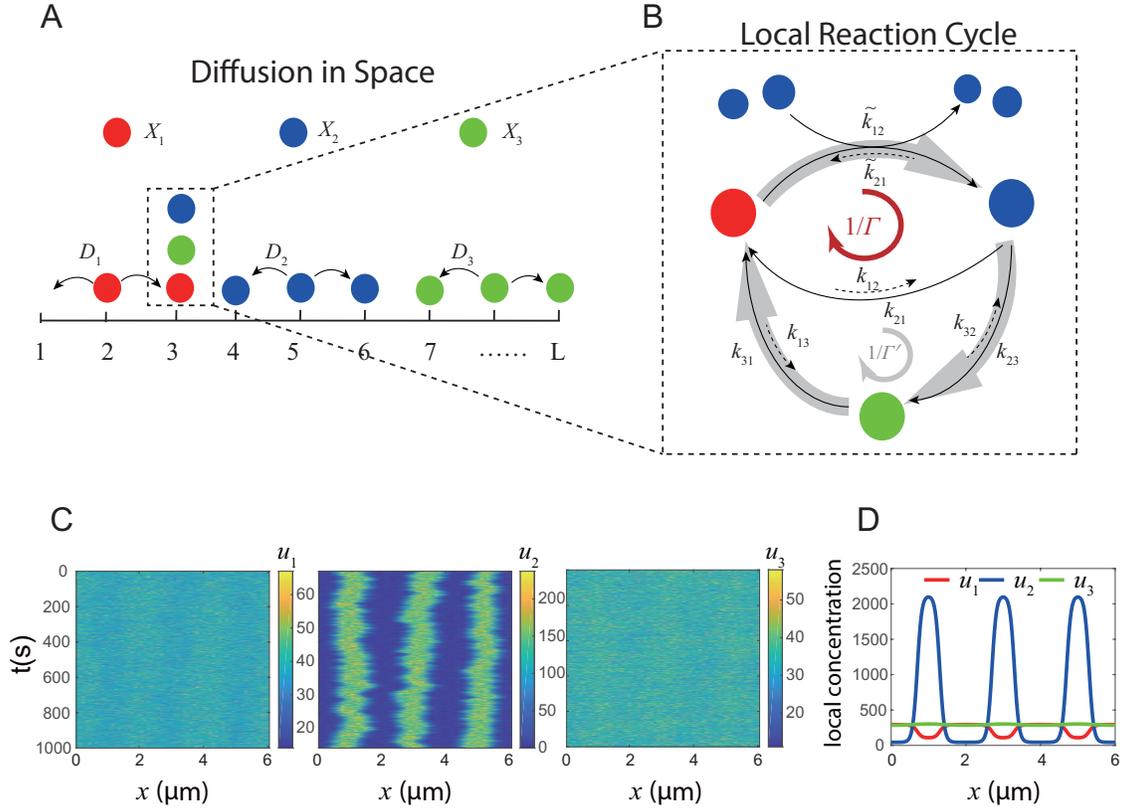}
   	\caption{ Schematic illustration of the stochastic reaction-diffusion system and its typical behavior. (A) Three different bio-molecules, $X_1$, $X_2$ and $X_3$,  represented by different colors diffuse in physical space with different diffusion constants $D_{1,2,3}$. (B) Within the same physical location (``box") (labeled by the ``box" number $i=1,2,...,L$) as shown in the dotted box in (A), the three type of molecules interact with each other through $4$ chemical reactions. These reactions form $2$ reaction cycles that are characterized by their irreversibility parameters $\Gamma$ and $\Gamma'$. (C) The spatial-temporal plots for the concentration fields $u_1(x,t)$, $u_2(x,t)$, and $u_3(x,t)$ for $X_1$, $X_2$, and $X_3$, respectively, in the Turing pattern regime. (D) The time averaged spatial profiles of $u_1$, $u_2$, and $u_3$.  
}
   	\label{fig:1}
   \end{figure}

The reactions given in Eq.~\ref{reaction} especially the autocatalytic reaction are similar to the Brusselator model for chemical oscillations. However, different from the well mixed systems, the molecules, $X_1$, $X_2$, and $X_3$, can diffuse with different diffusion constants $D_1$, $D_2$, and $D_3$, respectively. It was first shown by Turing in 1952~\cite{Turing1952} that when the reaction rates and the diffusion constants satisfy certain condition, the spatially homogeneous steady state will become unstable (Turing instability), and the system can spontaneously form spatially inhomogeneous pattern, which are now called the Turing pattern. One of the key requirement for the Turing pattern is that the diffusion constant of inhibitor is larger than that of activator:  
$d\equiv D_1/D_2> 1$
(we assume $D_1=D_3$ in this study).



This reaction-diffusion system can be considered as a nonequilibrium thermodynamic system~\cite{Rao2016,Falasco2018}, which can reach a nonequilibrium steady state (NESS) by continuously dissipating energy, e.g., by sustained reactant gradients in chemical reaction systems~\cite{Ouyang1991Transition} or continuous ATP hydrolysis in biological systems, which will be described later in this paper. In a typical biological system with a small number of molecules, there can be large fluctuations in the Turing pattern. In this paper, we focus on studying the relation between positional precision of the Turing pattern and the energy dissipation rate in a small reaction-diffusion system.



One of the main characteristics of nonequilibrium reaction networks is the existence of reaction cycles that carry persistent probability current even when the system reaches its steady state. There are two independent reaction cycles in the 3-state model: $X_1\to X_2\to X_1$ and $X_1\to X_2\to X_3\to X_1$ (see Fig.~1B for an illustration of the model).
The ratios of the products of the reaction rates in the counter-clock wise and clockwise in these two respective cycles are:
 \begin{equation}
\Gamma=\frac{\tilde{k}_{21} k_{12}}{\tilde{k}_{12} k_{21}},\;\;\;\;
\Gamma'= \frac{k_{13} k_{32} k_{21}}{k_{12} k_{23} k_{31}},
\label{gamma}
\end{equation}
which characterize the irreversibility of the two reaction cycles in the 3-node biochemical network as shown in Fig.~1B. The system is in equilibrium only when $\Gamma=\Gamma'=1$. When either of these two irreversibility parameters is different from $1$, the system is out of thermal equilibrium and energy is dissipated continuously even when the system is in its steady state. As the system is driven far from equilibrium, i.e., when $\Gamma$ is lower than a critical value, spatial homogeneity is spontaneously broken and Turing pattern emerges. A typical Turing pattern in our system and its time-averaged profiles are shown in Fig.~1C\&D. Next, we consider the energy cost of the reaction diffusion system underlying the Turing pattern.

\subsection{Dissipation in spatially extended reaction-diffusion systems}

For a spatially extended system, the free energy dissipation rate consists of two parts: the first part is due to local chemical reactions and the second corresponds energy dissipated to maintain nonuniform concentration field. Due to the spatial dependence of the concentration fields, we compute the energy dissipation rate per unit length for 1-D system studied here 
with the general definition of dissipation rate $\dot{W}_{individual}$ for each individual reaction~\cite{Qian2007}: 
\begin{equation}		\label{energy dissipation Qian}
\dot{W}_{individual}=(J^+-J^-)\ln(\frac{J^+}{J^-}),
\end{equation}
where $J^+$ and $J^-$ are forward and backward fluxes respectively between two microscopic states.

For the chemical reactions, the local dissipation rate {\it density} $\dot{w}(x)$ at position $x$ can be computed the same way as in homogeneous systems:
		\begin{equation}
		\dot{w}_\text{chem}(x)=\sum_{i=1}^{N_r} [j_{i}^{+}(x)- j_{i}^{-}(x)] \ln(\frac{j_{i}^{+}(x)}{j_{i}^{-}(x)}),
		\end{equation}
where $N_r=4$ is the number of reactions in the biochemical network and $j_{i}^{+}(x)$ and $j_{i}^{-}(x)$ are the forward and backward flux {\it densities} of the $i$-th reaction at position $x$, and free energy is in unit of thermal energy ($k_BT$). 
For the reaction between $X_2$ and $X_3$, we have: $j^+(x)=k_{23} u_2(x)$, $j^-(x)=k_{32}u_3(x)$, with $u_i(x)$ the local concentrations of molecules $X_i$. 
		 
The dissipation due to transport in space such as diffusion can be calculated by considering the spatial degrees of freedom as state variables. We divide the space into small boxes with size $\Delta x$, the dissipation rate of the free energy density $\dot{w}_{diff} (x)$ due to diffusion between neighboring boxes can be obtained by considering the diffusive transport fluxes as the forward and backward fluxes in the extended state-space. In particular, the forward and backward diffusive fluxes for the molecule $X_k$ are $J_{D,k}^+(x)=u_k(x)\Delta x \times \tilde{D}_k$ and $J_{D,k}^-(x)=u_k(x+\Delta x)\Delta x \times \tilde{D}_k$,
where $\tilde{D}_k$ is microscopic transition rate scaled from diffusion rate: $\tilde{D}_k\equiv D_k/\Delta x^2$.
Plugging in these two fluxes into Eq.~\ref{energy dissipation Qian}, we have:
\begin{equation}
\dot{w}_\text{diff}(x)= \sum_{k=1}^3 \frac{D_k(\frac{\partial u_k}{\partial x})^2}{u_k(x)},
\end{equation}
where $u_k(x)$ is the local concentration of $X_k$ molecule and the summation goes over all species $(k=1,2,3)$. 

The total energy dissipation rate $\dot{W}$ for the whole system is the sum of these two dissipation rate densities $\dot{w}_{chem}(x)$ and $\dot{w}_{diff}(x)$ integrated over space. In steady state, the net fluxes of reaction and diffusion  should balance each other for all molecule species. Using these steady state conditions, we can drastically simplify the expression for the total dissipation rate   (see SI for details):
\begin{align}
\dot{W}&=\int (\dot{w}_\text{diff}+\dot{w}_\text{chem})dx =\sum_{i=1} ^{N_r} ln(\frac{k_{i}^{+}}{k_{i}^{-}}) \int_{0}^{L} [j^+_{i}(x)-j_i^-(x)]dx \notag\\
&=J_{c1}\ln \Gamma^{-1}+J_{c2}\ln\Gamma'^{-1}, \label{total}
\end{align}
where $k_{i}^{+}$ and $k_{i}^{-}$ are the forward and backward reaction rate constants for the $i$-th chemical reaction, and $J_{1c}$ and $J_{2c}$ are the total fluxes in the two cycles with irreversible parameters $\Gamma$ and $\Gamma'$, which can be expressed as: $J_{1c}=\int_0^L (k_{12}u_1(x)-k_{21}u_2(x))dx=k_{12}N_{1}-k_{21}N_{2}$, $J_{2c}=\int_0^L (k_{23}u_2(x)-k_{32}u_3(x))dx=k_{23}N_{2}-k_{32}N_{3},$ 
where $N_{i}$ is the total number of $X_i$ molecules in the system. Plugging these expressions in Eq.~\ref{total}, we arrive at a simple equation for the expression for the total dissipation rate:
\begin{equation}
\dot{W}=(k_{12}N_1-k_{21}N_2)\ln(1/\Gamma)+(k_{23}N_2-k_{32}N_3)\ln(1/\Gamma').
\label{dotW}
\end{equation} 

It is surprising that the diffusion constants do not appear explicitly in the above expression (Eq.~\ref{dotW}) for the total energy dissipation. However, this result is intuitively reasonable since diffusion is not an active process and it only affects dissipation when chemical concentrations $u_k(x)$ and thus the fluxes, such as $J_{1c}$ and $J_{2c}$ become spatially non-uniform and additional energy is needed to overcome diffusion to maintain the spatial inhomogeniety in the Turing pattern.

\section{Results}

\subsection{Turing pattern and its free energy cost} 
 
In the 3-state model, the effect of $X_3$ and the $\Gamma'$ cycle is to localize the average position of the Turing stripes over a much longer time scale given that the $X_3$-related kinetic rates, i.e., $k_{13}$, $k_{23}$, $k_{32}$, and $k_{31}$ are much smaller than other rate constants. A detailed analysis on the role of $X_3$ is given in the SI (also see Murray and Sourjik~\cite{Murray2017} for a related discussion).  Overall, the energy cost and spatial precision of the Turing pattern are predominately controlled by the $\Gamma$ cycle. Therefore, we focus on studying the dependence of the dynamics and energetics of this biochemical network on $\Gamma$, which characterizes the chemical driving force in the system. In this study, we vary $\Gamma$ by changing $\tilde{k}_{21}$ while keeping other kinetic rates fixed, and define $W\equiv -\ln(\Gamma)$ to measure the dominant chemical driving force.    

	\begin{figure}
		\centering
		\includegraphics[width=0.9\linewidth]{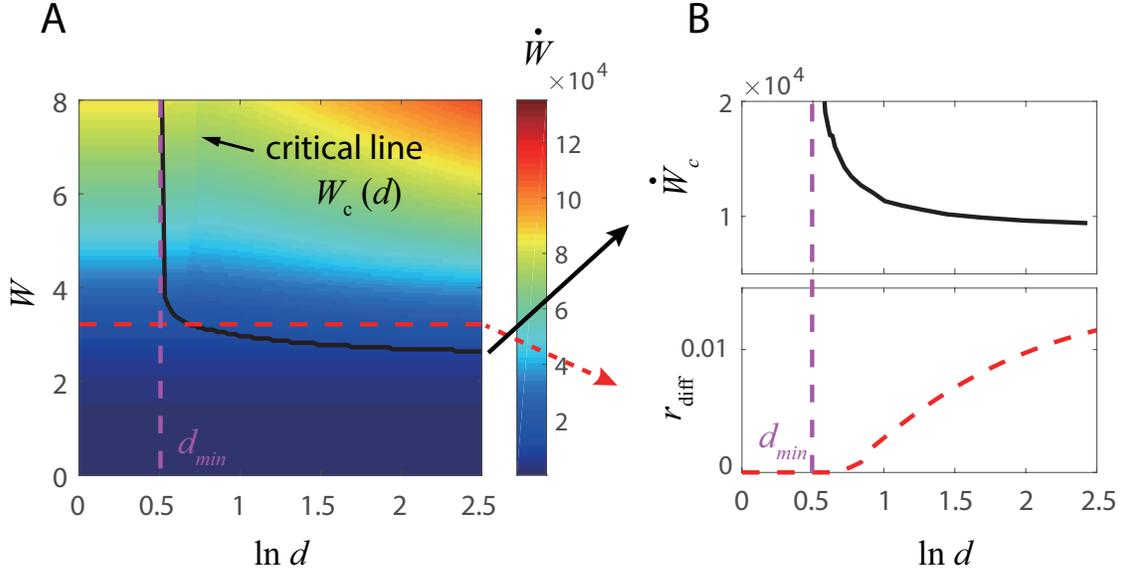}		
		\caption{Onset of the stochastic Turing pattern and its energy cost. 
		(A) The dependence of the energy dissipation rate ($\dot{W}$) on the chemical driving force ($W$) and the diffusion constant ratio ($d$). The black line represents the critical line ($W_c(d)$) for the onset of Turing pattern. The vertical purple dotted line shows the minimum value of $d$, $d_{min}$, below which no Turing pattern is possible independent of the chemical driving force $W$.  
(B) The critical energy dissipation rate $\dot{W}_c\equiv \dot{W}(W_c(d),d)$ (solid black line) versus $d$ When $d<d_{min}$, Turing pattern does not exist. For $d>d_{min}$, $\dot{W}_c$ decreases with $d$ but saturates to a finite value when $d\rightarrow \infty$. The red dotted line shows the fraction of energy dissipation due to diffusion $r_\text{diff}=\int \dot{w}_\text{diff}dx/\dot{W}$ versus $d$ for a fixed $W=3.21$, which corresponds to the red dotted line in (A). $r_\text{diff}=0$ before the onset of the pattern when $W<W_c(d)$, and increases with $d$ after the onset but saturate to a small value at $d\rightarrow \infty$. }
		\label{fig:2}
	\end{figure}

As first discovered by Turing, pattern formation also depends on the diffusion constant ratio $d$: only when $d$ is larger than a critical value $d_c$ the spatially homogeneous steady state can become unstable. Here, we study pattern formation and its energy dissipation rate ($\dot{W}$) in the parameter space spanned by the chemical driving force ($W$) and the diffusion constant ratio $d$.

In Fig.~\ref{fig:2}A, we show the dependence of energy dissipation rate $\dot{W}$ on the chemical driving force $W$ and $\ln d$. The transition from homogeneous state (no pattern) to a 3-stripe Turing pattern is shown by the solid line in Fig.~\ref{fig:2}A. We find that the onset of pattern formation occurs as the chemical driving force becomes larger than a critical value $W_c(d)$, which decreases with $d$. However, even in the limit $d\rightarrow\infty$, $W_c$ remains finite. 
The finite $W_c$ for all values of $d$ means a finite critical energy dissipation rate $\dot{W}_{c}$ is needed to generate and maintain the spatial organization (pattern). On the other hand, when $d$ is less than a minimum value $d_{min}\approx 1.7$, no pattern formation is possible even with an infinite chemical driving force as is shown in Fig.~\ref{fig:2}B. 


The overall dissipation rate consists of two parts: the dissipation in the chemical reactions and the dissipation used to overcome diffusion in order to maintain gradients. Here, we define $r_\text{diff}\equiv\int \dot{w}_\text{diff}dx/\int (\dot{w}_{diff}+\dot{w}_{chem})dx$ as the fraction of the energy dissipation used to overcome diffusion. Before the onset of Turing pattern, the concentration fields are spatially uniform and thus the dissipation is due to chemical reactions alone and $r_\text{diff}=0$. When $W > W_c(d)$, Turing pattern emerges, and $r_\text{diff}$ becomes nonzero. The dependence of $r_\text{diff}$ on $d$ with a fixed $W$ is shown in Fig.~\ref{fig:2}B (red dotted line). As expected, $r_\text{diff}$ generally increases with $d$ when $d>d_c(W)$ where $d_c(W)$ is the critical diffusion constant ratio at a given $W$. Note that $d_c(W)$ decreases with $W$ and it approaches $d_{min}$ when $W\rightarrow \infty$ (or $\Gamma=0$), i.e., $d_{min}=d_c(W=\infty)$ (the purple dotted line in Fig.~\ref{fig:2}A\&B). The ratio between the two dissipation rates can be estimated (see SI for details): $\frac{\int \dot{w}_{diff} dx}{\int \dot{w}_{chem} dx}\approx \frac{2}{\pi^2 W} (\frac{\Delta u_2}{\langle u_2\rangle})^2$ in the limit when $d\gg 1$ and $W\gg 1$, where $\Delta u_2 =(u_{2,max}-u_{2,min})/2$ is the amplitude of the spatial variation in $u_2(x)$ with $u_{2,max}$ and $u_{2,min}$ the maximum and minimum values of the concentration field $u_2(x)$ for molecule $X_2$,  and $\langle u_2\rangle$ the average of $u_2(x)$ over space (Note that we use $X_2$ because it shows the most significant spatial variation (pattern) among the three molecule species in our model). Overall, most of the energy is dissipated to drive chemical reactions to generate the Turing instability. After the onset of Turing pattern, the fraction of energy used for maintaining the spatial gradients against diffusion becomes non-zero and it increases as the relative amplitude of the Turing pattern increases. However, $r_{diff}$ remains to be small even deep in the Turing pattern regime. 



\subsection{The error-energy relation for Turing pattern in small systems} 

\begin{figure}
       	\centering
\includegraphics[width=1.0\linewidth]{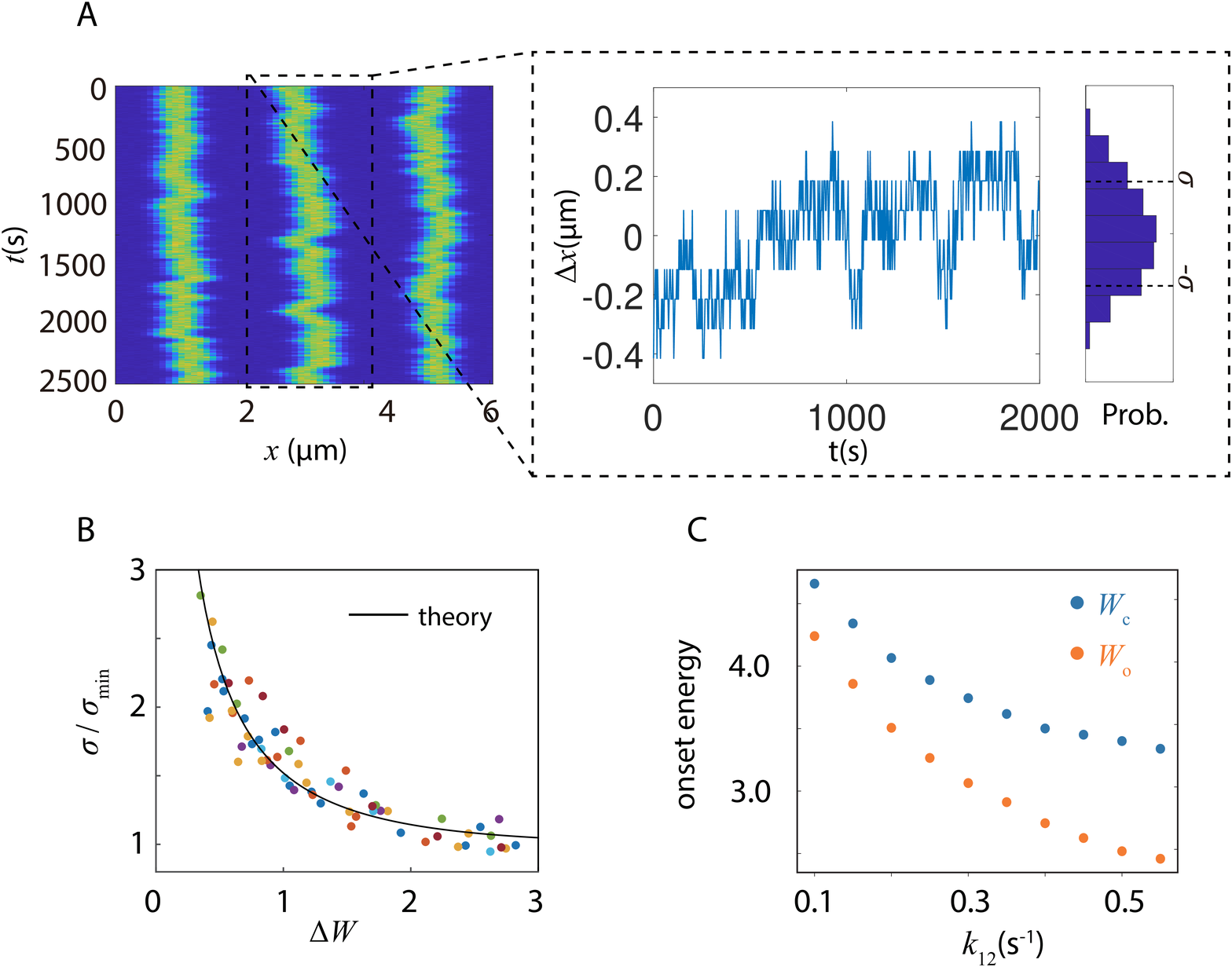} 
\caption{The accuracy-energy relationship in Turing pattern. 
(A) The spatial-temporal profile (kymograph) of $u_2(x,t)$. The fluctuations of the peak position $x_p(t)$ of the central stripe is shown in the dotted box. $\Delta x=x_p(t)-\left<x_p\right >_t$ is the deviation of the peak position from its mean at time $t$. The distribution of $\Delta x$ with a variance $\sigma$ is also shown. 
(B) The positional error $\sigma/\sigma_{min}$ versus the free-energy dissipation per cycle in additional to the critical energy, $\Delta W$. $\Delta W$ is varied by changing $\tilde{k}_{21}$ for different values of  $k_{12}=0.1,0.15,0.2,0.25,0.3,0.35,0.4,0.45,0.5,0.55\text{s}^{-1}$, which are represented by different colors. All data for different choices of $k_{12}$ and $\tilde{k}_{21}$ collapse onto the same curve that can be fitted by our theoretical prediction, Eq.~\ref{error_energy_1}, with fitted parameters: $c_1=0.93$ and $c_i=0$ for $i\ge 2$ (solid line). 
(C) The onset energy for finite system ($W_c$) and infinite system ($W_0$) versus $k_{12}$. It's clear that $W_c >W_0$ and both increases as $k_{12}$ decreases. Other parameters used are: $\tilde{k}_{12}=1.67\times 10^{-5}\text{s}^{-1}\mu\text{m}^{2}$, $k_{21}=3.6\text{s}^{-1}$, $k_{13}=k_{23}=0.0139\text{s}^{-1}$, $k_{31}=0.0416\text{s}^{-1}$, $k_{32}=1.39\times 10^{-5}\text{s}^{-1}$, $D_1=D_3=0.3\mu\text{m}^{2}\text{s}^{-1}$, $D_{2}=0.012\mu\text{m}^{2}\text{s}^{-1}$.}
       	\label{fig:3}
       \end{figure}

Turing patterns spontaneously break the spatial translation symmetry of the underlying homogeneous biochemical reaction-diffusion system. As a result, the ``phase" degree of freedom of the Turing pattern is a soft mode that can have large fluctuations due to noise in finite biochemical systems with a small number of molecules. In Fig.~\ref{fig:3}A, a time series of the peak location ($x_{p}(t)$) for one of the molecular species $X_2$ is shown. The standard deviation ($\sigma$) of $x_{p}$, 	 
$\sigma\equiv  \sqrt{\langle(x_{p}(t)-\bar{x}_{p})^{2}\rangle_t}$, can be used as a measure of the spatial error of the Turing pattern. 

In an infinite system, the most unstable mode has a wavevector $q_0$, which is the wavevector with the highest linear growth rate $\rho(q)$, i.e., $\frac{d \rho}{d q}|_{q_0}=0$.  In a finite system with size $L$, the Turing pattern wavevector $q_n=\frac{2\pi}{\lambda}$ where $\lambda=L/n$ with $n\ge 1$ the integer wavenumber. Typically, $q_n\ne q_0$ and their difference is given by $\Delta q \equiv q_n -q_0(\ne 0)$. The spatial-temporal profile of a concentration field (e.g., $u_2(x,t)$) in a Turing pattern can be written as: $u_2(x,t)=\Theta (q_n x +\phi(x,t))$, where $\Theta$ is a periodic function with period $2\pi$ and $\phi(x,t)$ is the phase variable of the Turing pattern.  The phase variable satisfied the phase diffusion equation~\citep{Cross1993Pattern}, which can be generally written as :
\begin{equation}
\frac{\partial \phi}{\partial t}=D_2 \frac{\partial^2 \phi}{\partial x ^2}-D_4 \frac{\partial^4 \phi}{\partial x ^4} + \partial_x \eta,
\label{ph_eq}
\end{equation}
where $D_2$ is the second order diffusion term and a 4th order diffusion term with $D_4>0$ is introduced to prevent divergence when $D_2\rightarrow 0$. The form of the noise term $\partial_x \eta$ is due to the translational invariance of the phase variable $\phi$ and $\eta$ is a Gaussian white noise: $\langle \eta(x,t)\eta(x',t')\rangle =\Delta_0 \delta(t-t')\delta(x-x')$ with $\Delta_0$ the noise strength.

Following the standard procedure~\cite{Cross1993Pattern}, the second order phase diffusion constant $D_2$ can be expressed as:
      \begin{equation}
      D_2 =\xi ^2\tau^{-1}\frac{\epsilon_0 -3\xi^2 \Delta q^2}{\epsilon_0 -\xi^2 \Delta q^2},
      \label{D2}
\end{equation}
where the control parameter is defined as $\epsilon_0 \equiv 1- \tilde{k}_{21}/\tilde{k}_0$ with $\tilde{k}_0$ the critical value of $\tilde{k}_{21}$ in an infinite system, $\xi$ is a characteristic length given by: $\xi^2 =-\frac{1}{2}\frac{d^2\rho}{dq^2}|_{q_0}$, and $\tau$ is a characteristic timescale. In an infinite system, Turing pattern appears when $\epsilon_0\ge 0$ or equivalently $\tilde{k}_{21}\le \tilde{k}_0$. 
In a finite system, however, the requirement for a Turing pattern with phase stability ($D_2>0$) becomes more stringent due to a non-zero $\Delta q( \ne 0)$.  
The critical value $\tilde{k}_c$ for a stable Turing pattern with wavevector $q_n$ in a finite system can be defined as the value of $\tilde{k}_{21}$ when the phase diffusion constant becomes zero, i.e., $D_2=0$. From Eq.~\ref{D2}, we determine $\tilde{k}_c=\tilde{k}_0 (1-3\xi^2 \Delta q ^2)<\tilde{k}_0$, which represents a stronger requirement than that in the infinite system. Based on this critical value $\tilde{k}_c$, a new control parameter can be defined as: $\epsilon \equiv 1- \tilde{k}_{21}/\tilde{k}_c$ for a finite system. The phase diffusion constant $D_2(\epsilon)$ is an increasing function of $\epsilon$ with $D_2(\epsilon=0)=0$. 
From $\tilde{k}_c$, we can also define a critical value $\Gamma_c \equiv \frac{k_{12}\tilde{k}_{21}}{k_{21}\tilde{k}_c}$ for $\Gamma$, so $\epsilon =1-\Gamma/\Gamma_c$. 
Note that $\Gamma_c <\Gamma_0$ because $\tilde{k}_c<\tilde{k}_0$, which means the phase stability of the Turing pattern in a finite system requires a higher dissipation rate (per molecule) than the onset energy $W_0(\equiv -ln(\Gamma_0)$ in the infinite system. 

From the stochastic phase equation (Eq.~\ref{ph_eq}), we can compute the positional variance $\sigma^2$, which is proportional to the phase variance: 
         \begin{equation}\sigma^2 \equiv  (\frac{\lambda}{2\pi})^2 \langle \phi^2\rangle 
=  (\frac{\lambda}{2\pi})^2 \int_{\pi/L}^{\infty}\int_{-\infty}^{\infty} \frac{\Delta_0 q^2 d\omega dq}{\omega^2+(D_2(\epsilon)q^2+D_4q^4)^2} =\frac{\sigma_0^2}{S(\epsilon)},
         \label{phase}
         \end{equation}
where $\omega$ and $q$ represent the frequency and wave vector  respectively; $\sigma_0^2$ is the position variance when $\epsilon=0$ (or $\Gamma=\Gamma_c$) and $S(\epsilon)$ is the variance reduction factor, which is an increasing function of $\epsilon$ with $S(0)=1$. Given that $D_2=0$ when $\epsilon=0$, we can use a linear approximation for $D_2=d_2\epsilon$ with a positive constant $d_2(>0)$. By further assuming a constant $D_4>0$ in Eq.~\ref{phase}, we have $\sigma_0^2 = (\frac{\lambda}{2\pi})^2 \Delta_0 L d_2 D_4$ and $S(\epsilon)=a \epsilon^{1/2} /tan^{-1}(a\epsilon^{1/2})$ with a constant $a=\frac{L}{\pi}(d_2/D_4)^{1/2}$. 

Eq.~\ref{phase} clearly shows that the positional error $\sigma$ decreases as $\epsilon=1-\tilde{k}_{21}/\tilde{k}_n=1-\Gamma/\Gamma_c$ increases or equivalently when $\Gamma$ decreases.  
 According to Eq.~\ref{total}, the total energy dissipation can be decomposed into those from each cycle: $\dot {W} = \dot{W}_1+\dot{W}_2=J_{1c} \ln(\Gamma^{-1})  +J_{2c} \ln(\Gamma'^{-1}) $, where $J_{1c}$ and $J_{2c}$ are the fluxes of the two cycles integrated over space. In the 3-state model, the energy dissipation is dominated by the first cycle as the flux in the first cycle is much larger than that in the second cycle: $J_{1c}\gg J_{2c}$ or equivalently the cycle time $\tau_1\equiv J_{1c}^{-1}$ for the first cycle is much shorter than that of the second cycle $\tau_2\equiv J_{2c}^{-1}$: $\tau_1 \ll \tau_2$. 
As a result, the total energy dissipation per molecule during the dominant cycle time $\tau_1$ is: $ \ln(\Gamma^{-1}) +\frac{\tau_1}{\tau_2} \ln(\Gamma'^{-1})\approx  \ln(\Gamma^{-1})=W$, which is approximately the chemical driving force defined before. Let $W_c \equiv \ln(\Gamma_c^{-1})$ denote the critical (onset) energy dissipation per cycle in the finite system, we have $\epsilon =1-\exp(- \Delta W)$ where $\Delta W \equiv  W- W_c$ is the additional energy dissipation per cycle beyond the critical energy dissipation $ W_c$. 


In general, the system is most stable in the limit of $\epsilon \rightarrow 1$, i.e., the strong driving limit $\Delta W \rightarrow \infty$, where the error $\sigma(\epsilon=1)\equiv \sigma_{min}= \sigma_0/\sqrt{S(1)}$ is at its minimum. From Eq.~\ref{phase}, we can write $\sigma (\epsilon) =\sigma_{min}/r(\epsilon)$ with an error reduction function $r (\epsilon)\equiv (S(\epsilon)/S(1))^{1/2}$. Since $S(\epsilon)$ is an increasing function of $\epsilon$, $r(\epsilon)$ is also an increasing function of $\epsilon$ with $r(1)=1$.  In the strong driving (or high dissipation) limit, we can expand $r(\epsilon)$ around $\epsilon=1$:  $r(\epsilon)=1+\sum_{i=1}c_i (\epsilon-1)^i$ with constant coefficients $c_i$ ($c_1>0$). From Eq.~\ref{phase} and by using the dependence of $\epsilon$ on $\Delta W$, we obtain the error-energy relation: 
       \begin{equation}
	\sigma= \frac{\sigma_{min}}{r(\epsilon)} = \frac{\sigma_{min}}{1-c_1 \exp(-\Delta W)+h.o.t. },
	\label{error_energy_1}
	\end{equation} 
where only the first leading order terms ($c_1$) is written out explicitly for simplicity ($h.o.t.$ stands for higher order terms). Eq.~\ref{error_energy_1} clearly shows that positional error can be suppressed by increasing energy dissipation.  

We have tested this error-energy dependence (Eq.~\ref{error_energy_1}) by extensive simulations of the 3-state model for different values of $\tilde{k}_{21}$ and $k_{12}$. As shown in Fig.~\ref{fig:3}B, the dependence of the normalized positional error $\sigma/\sigma_{min}$ on the additional energy dissipation $\Delta W$ collapsed onto the same curve that can be fitted by Eq.~\ref{error_energy_1} for all different values of $\tilde{k}_{21}$ and $k_{12}$. The dependence of $W_{0}$ and $ W_c$ on $k_{12}$ are shown in Fig.~\ref{fig:3}C, which clearly shows that the critical energy for a finite system is larger than that for the infinite system: $W_c>W_0$ and both decrease with $k_{12}$. Both the analytical and numerical results clearly show that a larger dissipation per cycle $\Delta W$ (or equivalently a smaller $\Gamma$) suppresses the phase fluctuations and leads to a higher positional accuracy in Turing pattern. In a recent study~\cite{Barato2020}, a non-monotonic dependence of error on dissipation was found in a 2-state model with periodic boundary condition. The increase of spatial error in the large dissipation limit may be caused by the existence of multiple metastable patterns in the simple 2-state model with periodic boundary conditions. In the 3-state model studied here, we do not observe the increase in positional error as dissipation increases, likely due to the effect of the additional molecular species $X_3$ in localizing the average position of the Turing pattern and the realistic boundary condition~\cite{Murray2017}, which also serves to suppress the metastable states  . 

\subsection{Free energy dissipation enhances the robustness of Turing pattern against concentration fluctuations}

\begin{figure}
\includegraphics[width=1.0\linewidth]{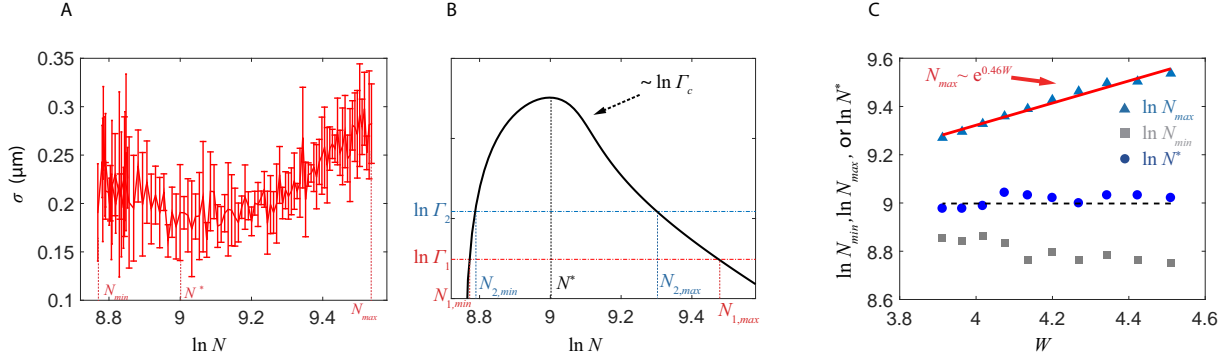} 
\caption{Dependence of Turing pattern on total molecule number. 
(A) Positional error $\sigma$ has a U-shape dependence on the  molecule number $N$. The 3-stripe Turing pattern is stable in a finite range $N_{min}\le N\le N_{max}$ with $\sigma$ reaching its minimum at $N^*$. 
(B) The non-monotonic dependence of $\Gamma_c$ on $N$. 
Two different choice of $\Gamma$ are shown to illustrate how $N_{min}, N_{max}$ and $N^*$ should vary with $\Gamma$.
 (C) The dependence of $N_{min}, N_{max}$ and $N^*$ on $W$ (energy dissipation per cycle). A linear fit between $\ln N_{max}$ and $W$ (red line) has a slope $0.46$, which is close to the theoretical value $0.5$ (Eq.~\ref{Nmax}). $N^*$ is independent of $W$ and close to the maximum position of $\Gamma_c(N)$ (black dotted line). 
Parameters used here are: $\Gamma=0.011$, $\tilde{k}_{21}=1.67\times 10^{-5}\text{s}^{-1}\mu\text{m}^{2}$,$k_{12}=0.5\text{s}^{-1}$, $k_{21}=3.6\text{s}^{-1}$, $k_{13}=k_{23}=0.139\text{s}^{-1}$, $k_{31}=0.416\text{s}^{-1}$, $k_{32}=0.0139\text{s}^{-1}$, $D_1=D_3=1.8\mu\text{m}^{2}\text{s}^{-1}$, $D_{2}=0.012\mu\text{m}^{2}\text{s}^{-1}.$}
       	\label{fig:4}
       \end{figure}

In small biological systems such as a cell, protein concentration can fluctuate in time and vary from cell to cell~\cite{Elowitz2002,rao2002control,Xie2010,Raser2004,Salman2012Universal,Sassi2022Protein}. Here, we study how the positional error $\sigma$ depends on the molecule (protein) concentration by varying the total molecule number $N$ in the 3-state model with a fixed length $L$. Intuitively, since the overall noise level (fluctuation) scales as $N^{-1/2}$, increasing $N$ is expected to lead to a higher spatial accuracy. 
However, in a biochemical reaction system with nonlinear reaction dynamics, increasing $N$ also affects the system's sensitivity to fluctuations, which makes the overall effect of varying $N$ on spatial accuracy unclear. From our simulation results, we found that a specific spatial pattern (e.g., a stable 3-stripe pattern) only exists in a finite range of molecule copy number: $N_{min}<N<N_{max}$. The system transitions to other spatial patterns (e.g., 2-stripe patterns) when $N$ is outside of this range. As shown in Fig.~\ref{fig:4}A, the dependence of $\sigma$ on $N$ (for $N_{min}<N<N_{max}$)  follows a non-monotonic ``U"-shape and there exists an optimal molecule number $N^*$ where the positional error  $\sigma$ is minimal. 
More specifically, $\sigma$ does decrease as $N$ increases when $N_{min}<N<N^*$, however, for $N^*<N<N_{max}$, the positional error $\sigma$ increases as $N$ increases, which is counter intuitive. 


How does this non-monotonic dependence of $\sigma$ on $N$ arise? As shown before, the positional error of the Turing pattern can be written as 
$\sigma=\sigma_{min}r^{-1}(\epsilon)\propto\Delta^{1/2}r^{-1}(\epsilon)$ where the overall noise intensity is inversely proportional to the total molecule number $\Delta\propto N^{-1}$ and the inverse noise reduction factor $r^{-1}(\epsilon)$ can be understood as the sensitivity (susceptibility) to noise. To understand the $N$-dependence, we define the relative concentrations $v_i\equiv u_i/c_{tot}$ ($i=1,2,3$) with $c_{tot}=N/L$ the total molecule concentration.  Dynamics of the relative concentrations $v_i$ are governed by the same equations as those for $u_i$ but with effective reaction rates. For the linear reactions, the effective reaction rates remain the same as the original rates. However, for the nonlinear reactions, e.g., the autocatalytic reaction, the effective reaction rates are normalized by $c_{tot}$: $\beta_{12(21)}=\tilde{k}_{12(21)} c_{tot}^2=\tilde{k}_{12(21)} N^2/L^2$. Thus, the noise susceptibility $r^{-1}(\epsilon)$ depends on $N$ because the control parameter $\epsilon\equiv1-\Gamma/\Gamma_c$ depends on the critical value $\Gamma_c$, which depends on $N$ through its dependence on $\beta_{12}$ and $\beta_{21}$. 


As described earlier in this paper, the critical reaction rate ($\tilde{k}_c$) for a stable 3-stripe pattern is proportional to the onset reaction rate ($\tilde{k}_0$) in an infinite system:  $\tilde{k}_c=\zeta\tilde{k}_0 $ with $\zeta=1-3\xi^2 \Delta q ^2$ approximately a constant. Therefore, we have
$\Gamma_c\equiv \frac{\tilde{k}_ck_{12}}{\tilde{k}_{12}k_{21}} =\zeta \frac{\tilde{k}_0k_{12}}{\tilde{k}_{12}k_{21}}  \equiv \zeta \Gamma_0$ where $\Gamma_0$ can be expressed as: 
\begin{equation}\label{onset0}
\Gamma_0\equiv\frac{\beta_{0}k_{12}}{\beta_{12}k_{21}},
\end{equation}
with  
$\beta_{0}$ the critical effective rate of $\beta_{21}$, which can be determined analytically by the linear stability analysis of the dynamic equations for $v_i$ (see SI for details).  In the limit $d\gg 1$, we have:: 
\begin{equation}\label{onset}
\beta_{0}=-2\frac{k_{12}}{R_1}\frac{1}{v_2^{*3}}+\left(2k_{12}\frac{R_2}{R_1}+k_{21}\right)\frac{1}{v_2^{*2}},
\end{equation}
where 
$
R_1=(k_{31}+k_{32}+k_{13})/({k_{31}+k_{32}})>1$ and $
R_2=(k_{31}+k_{32}+k_{23})/({k_{31}+k_{32}})>1
$ are two constants,  
and $v_{2}^*$ is the relative concentration of the homogeneous fixed point solution, which depends on $N$ via its dependence on $\beta_{12}$ and $\beta_{21}$.
As $N$ increases, the nonlinear autocatalytic  reaction becomes more dominant as both $\beta_{12}$ and $\beta_{21}$ increase with $N$. For typical kinetic rates with $k_{12}\ll k_{21}$ and $\tilde{k}_{12}\gg\tilde{k}_{21}$ as used in our model, the dominance of the autocatalytic reaction at larger $N$ leads to a higher value of $v_2^*$, i.e., $v_2^*$ increases with $N$.
Finally, since the cubic and quadratic terms in Eq.~\ref{onset} have opposite signs,  $\beta_{0}$ and therefore $\Gamma_0$ can be a non-monotonic function of $v_2^*$ and consequently a non-monotonic function of $N$.  

 
By using Eqs.~\ref{onset0}\&\ref{onset}, we can compute the dependence of $\Gamma_c(N)=\zeta\Gamma_0(N)$ on $N$ numerically. As shown in Fig.~\ref{fig:4}B, as $N$ increases, $\Gamma_c$ first rises sharply to a peak at $N^*$ before decreasing more gradually. 
For given values of reaction rates, the range of $N$ for the 3-stripe Turing pattern is set by $\Gamma_c(N)=\Gamma(\equiv \frac{\tilde{k}_{21} k_{12}}{\tilde{k}_{12} k_{21}})$, which determines the minimum and maximum molecule number $N_{min}$ and  $N_{max}$. The non-monotonic dependence of $\Gamma_c(N)$ on $N$ explains the origin of the U-shaped dependence of the positional error ($\sigma$) on the total number of molecules ($N$) and the finite range of $N$ for the existence of the Turing pattern in a system with a fixed size as observed in Fig.~\ref{fig:4}A.

It is clear from our analysis and Fig.~\ref{fig:4}B that both $N_{min}$ and $N_{max}$ change with $\Gamma$ or equivalently the energy dissipation of the system $W\equiv-\ln\Gamma$, whereas the optimal molecule number $N^*$ is independent of $W$. 
With an increased dissipation $W$ (by decreasing $\Gamma$), $N_{min}$ decreases and $N_{max}$ increases, both of which broaden the range defined by $R\equiv N_{max}/N_{min}$. 
Since the dependence of $\Gamma_c(N)$ on $N$ has a sharp rise and a more gradual decay, $N_{max}$ is more sensitive to the change of $W$ than $N_{min}$. 
To test this result, we determined $N_{max}$, $N_{min}$ and $N^*$ in our simulations for different values of $\Gamma$ or $W$.  As shown in Fig.~\ref{fig:4}C, $N_{max}$ increases with $W$ whereas $N_{min}$ decreases albeit weakly with $W$. $N^*$ almost keeps constant near the maximum position of $\Gamma_c(N)$.
In the limit of large dissipation when $\Gamma\ll 1$, $v_2^*$ will saturate for large $N$ and so will $\beta_0$, 
so the dependence of $\Gamma_c$ on $N$ is dominated by the factor $\beta_{12}^{-1}\propto N^{-2}$. 
As a result, we have 
(see SI for details):
$
\Gamma_c\approx\tilde\alpha N^{-2},
$
where $\tilde\alpha$ is a coefficient depending on model parameters. By using this asymptotic behavior of $\Gamma_c(N)$, we can solve  $\Gamma_c(N_{max})=\Gamma$ and obtain:
\begin{equation}\label{Nmax}
N_{max}\approx\left(\frac{\tilde\alpha}{\Gamma}\right)^{1/2}
\propto e^{W/2}.
\end{equation} 
Eq.~\ref{Nmax} shows that $N_{max}$ increases with $W$ exponentially, which is confirmed by numerical results shown in Fig.~\ref{fig:4}C. The steep increase of $\Gamma_c$ near $N_{min}$ indicates a relatively weak decrease of $N_{min}$ with $W$, which is also consistent with numerical results shown in Fig.~\ref{fig:4}C. Note that \eqref{Nmax} is derived under the condition $d\ll1$. 
For a finite $d$ and very small $\Gamma\ll d^{-1}$, $N_{max}$ saturates to a value controlled by $d^{-1}$ (see SI for detailed discussion). 

Overall, our results show that there is a finite range of concentrations over which a specific Turing pattern is stable due to nonlinearity in the reaction kinetics. A higher dissipation $W$ can broaden this range, which enhances the robustness of the desired Turing pattern against the inevitable concentration variations in living cells. 

\subsection{A realistic biological system}
Finally, we study the role of energy dissipation by considering a realistic biochemical system that achieves spatial positioning via the Turing mechanism, namely the Muk system~\cite{MukBEFScience2012, Murray2017} responsible for DNA segregation in {\it E.coli}.

A MukBEF complex consists of three kinds of proteins: a MukB dimer, which is a distant relative of Structural Maintenance of Chromosomes (SMC) protein family and is the core of the MukBEF complex, and two small accessory proteins MukE and MukF. The MukB dimer has a rod-and-hinge structure, which forms a loop to capture DNA. It also has an ATP binding domain, and experiments show that the MukB dimer serves as an ATP-dependent ``DNA binding switch'': ATP binding promotes attachment of the MukBEF complex to DNA whereas hydrolysis of the bound ATP stimulates DNA detachment. 

       \begin{figure}
       	\centering
       	\includegraphics[width=0.8\linewidth]{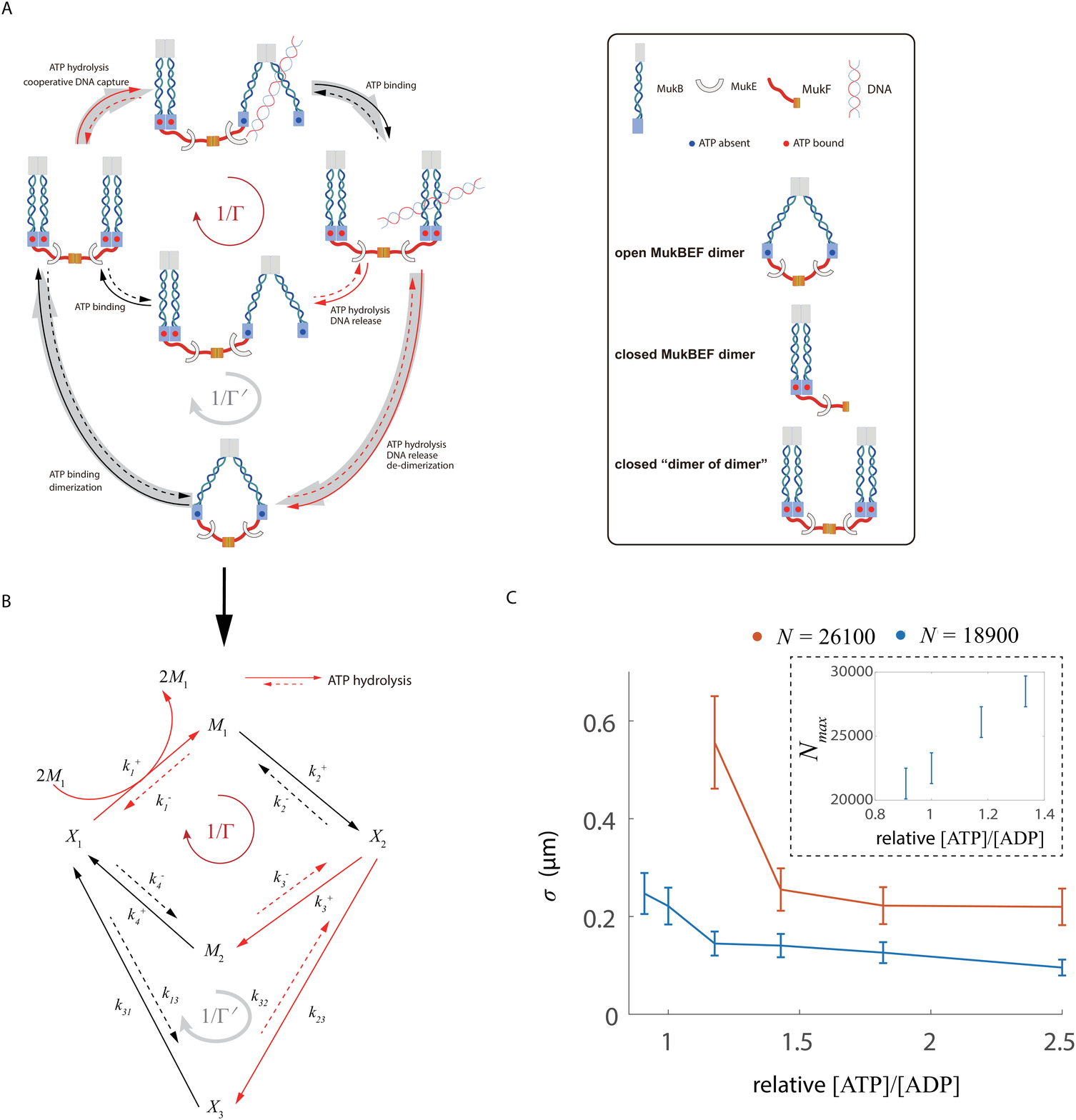}
       	       	\caption{A model of Muk system for DNA segregation. (A) Illustration of the Muk system. There are two cycles (red and grey) in our system with the red cycle controlling the pattern's positional order.
(B) The corresponding reaction network. Red lines indicate the reactions involving ATP hydrolysis. The ratio of the forward reaction rate and the backward reaction rate of these ATP-hydrolysis driven reactions is proportional to the $[ATP]/[ADP]$ concentration ratio.
(C) The dependence of the positional error $\sigma$ on the  $[ATP]/[ADP]$ concentration ratio for different total protein copy number $N$. Each data point is obtained by averaging 5 Gillespie simulations. At a given $[ATP]/[ADP]$ ratio, the Turing pattern disappears when $N\ge N_{max}$. The inset shows that the maximum protein copy number $N_{max}$ increases with the $[ATP]/[ADP]$ ratio. 
Details of the model and the parameters used are given in the SI.}
       	\label{fig:fig5}
       \end{figure}

The accurate spatial clustering of DNA-attached MukBEF complex is critical for chromosome organization. Based on functional and structural studies, a simple ``rock-climber'' model was proposed in \cite{MukBEFScience2012} to explain the working mechanism of the MukBEF complex, which is used here to study the role of energy dissipation in pattern formation. As illustrated in 
Fig.\ref{fig:fig5}A,
without binding to ATP, a MukB dimer remains at its ``open'' conformation, which cannot attach to DNA. Once an open MukB dimer bind with ATP, the MukBEF dimer transforms to a ``close'' conformation, which can dimerize with another closed MukBEF dimer to form a ``dimer of dimer'' (DD). 
A DD can attach to DNA by hydrolyzing ATP in one of its dimers, which leads to a conformational change from the close state to the open state in that dimer, which enables it to capture DNA. 
This capturing process is highly cooperative~\cite{Cui2008}, i.e., it is enhanced by having other MukB DD's nearby. 
Once the MukB dimer captures DNA, the attachment to DNA becomes tighter when it binds to ATP and returns to the close conformation. 
Once bound to DNA, the MukBEF DD becomes relatively immobile, i.e., the diffusion of the DNA bound MukB DD is much slower. 
However, a MukB dimer that is attached to DNA can become detached from DNA by hydrolyzing its bound ATP, which changes the dimer to its open conformation and releases DNA, and the next cycle is ready to start. In case when both ATP molecules bound to the DD are hydrolyzed (almost) simultaneously, besides releasing the attached DNA, the DD can also de-dimerize to form two separate open dimers. These open dimers have to bind with ATP and dimerize to become functional again. 

This reaction network can be simplified to a two-loop reaction-diffusion network similar to the network introduced in previous sections, as shown in Fig.\ref{fig:fig5}B. $X_2$ and $X_1$ represent the closed DD that are bound to the DNA or not, respectively; and $X_3$ represents the open MukBEF dimer. To describe the effects of ATP hydrolysis, we introduce two intermediate states $M_1$ and $M_2$ right after each ATP hydrolysis reaction in the DD loop ($X_1\rightarrow M_1\rightarrow X_2\rightarrow M_2\rightarrow X_1$). The red lines in Fig.\ref{fig:fig5}B represent all the reactions that are driven by ATP hydrolysis. For these ATP hydrolysis driven reactions, the ratio of the forward reaction rate and the backward reaction rate is proportional to the ATP/ADP ratio, e.g., $k_1^+/k_1^-\propto [ATP]/[ADP]$. The DNA free states $X_1$, $X_3$, and $M_2$ are assumed to have the same faster diffusion constant, whereas the DNA bound states $M_1$ and $X_2$ are assumed to have the same slower diffusion constant. See SI for the detailed description of the model.

We studied behaviors of this model of the Muk system for different $[ATP]/[ADP]$ ratio, which serves as a proxy for energy dissipation rate in the system. As shown in Fig.~\ref{fig:fig5}C, Turing pattern emerges in a wide range of $[ATP]/[ADP]$ ratio and the positional error $\sigma$ of Turing pattern decreases when the $[ATP]/[ADP]$ ratio increases. Furthermore, for a higher $[ATP]/[ADP]$ ratio, the pattern is more robust to variations in total MukB protein copy number. In particular, the largest MukB protein copy number $N_{max}$, below which Turing pattern is stable, increases with the $[ATP]/[ADP]$ ratio (see inset of Fig.~\ref{fig:fig5}C). These general predictions on the dependence of precision and robustness of Turing pattern on energy dissipation may be tested in future experiments by varying the ATP/ADP ratio in the system. 


\section{Conclusion and Discussion}   
 
Accurate spatial organization is critical for many biological processes and functions. However, spatial patterns can fluctuate and even become unstable due to strong noise in small biological systems. In this paper, we investigated whether and how energy dissipation in the underlying non-equilibrium reaction-diffusion systems is related to accuracy and robustness of the spatial pattern by studying a generic 3-state reaction-diffusion model motivated by realistic biological systems. We showed that there is a critical (minimum) energy cost ($W_c$) to create and maintain a Turing pattern and $W_c$ decreases as the ratio of the diffusion constants ($d$) increases and it saturates to a finite value as $d\rightarrow \infty$. As the energy cost increases beyond $W_c$, the spatial accuracy of the Turing pattern increases. A general trade-off relation (Eq.~\ref{error_energy_1}) between spatial error $\sigma$ and the energy cost is obtained by analyzing phase dynamics of the spatial pattern. In a finite system, 
we found that the positional error has a distinctive U-shape dependence on total molecule number $N$ and the Turing pattern is stable only in a finite range of $N$. A higher dissipation leads to a wider range of $N$ over which the spatial pattern is stable and thus enhances the robustness of the Turing pattern against biologically realistic molecule number variations. We have used this theoretical framework to study the MukBEF system responsible for DNA segregation in {\it E. coli}. Consistent with the general theoretical results, we found that the Turing pattern becomes more accurate and it exists in a wider range of $N$ as the ATP/ADP ratio increases, both of which can be tested in future experiments.  

In Turing patterns, spatial regularity arises in a homogeneous system based on an elegant reaction-diffusion (RD) mechanism that depends on the interplay between nonlinear activator-inhibitor chemical reactions and the different diffusion constants for the activator and inhibitor species in the system. There is, however, another class of more direct mechanisms for pattern formation based on preexisting asymmetry in the system, e.g., a sustained chemical gradient(s) across the entire length of the system. The representative model is the positional information (PI) model (aka the french flag model) first proposed by Wolpert~\cite{WOLPERT1969}, which has been verified in developmental pattern formation in {\it Drosophila}~\cite{Akam1989Making} and other organisms. These two mechanisms of pattern formation are obviously quite different and they apply to different biological systems (see \cite{Sharpe2015Positional} for a recent review). These two mechanisms are also different in terms of their energy cost. In the RD mechanism, the total number of molecules is conserved, and we showed that most of the energy is spent on driving the chemical reaction cycles that convert the molecules from one form to another, which gives rise to the pattern formation. The fraction of energy cost used to overcome diffusion for maintaining the spatial gradient is small. On the other hand, for the PI mechanism, the morphogene molecules have a finite life time, and most of the energy is used for synthesizing the morphogen molecules for maintaining the morphogene gradient. In particular, the localized synthesis of the morphogene protein molecule and its global degradation lead to a sustained morphogene gradient, which provides the positional information that can be read off by a down stream mechanism for pattern formation. However, despite the differences between the two mechanisms, as recently reported by Song and Hyeon~\cite{Song2021Cost}, there is an accuracy-cost trade-off relation in the PI mechanism, which is similar to what we found for the RD mechanism. This raises the question whether there is an universal relation between energy cost and accuracy in pattern formation systems independent of details of the underlying mechanisms, which may provide an interesting direction for future study. In general, we believe the theoretical framework based on nonequilibrium thermodynamics provides a novel lens for investigating biological systems in search of possible unifying principles. 




\section{Acknowledgments}
We thank Lei Zhang for useful discussions. The work of DZ and QO is supported by NSFC (12090054), and DZ also acknowledges support form China Postdoctoral Science Foundation (2020M680180). The work by YT is partially supported by NIH (R35GM131734).

\newpage
\bibliographystyle{unsrt}
\bibliography{ref.bib}

\end{document}


\title{Supplementary Information on ``The error-energy relation and optimal molecule copy number for Turing patterns in finite spatially extended systems'' }

\author{Dongliang Zhang}
\affiliation{The State Key Laboratory for Artificial Microstructures and Mesoscopic Physics, School of Physics, Peking University, Beijing 100871, China}
\author{Chenghao Zhang}
\affiliation{The State Key Laboratory for Artificial Microstructures and Mesoscopic Physics, School of Physics, Peking University, Beijing 100871, China}
\affiliation{Physics Department, University of Illinois, Urbana, IL61801}
\author{Qi Ouyang}
\affiliation{The State Key Laboratory for Artificial Microstructures and Mesoscopic Physics, School of Physics, Peking University, Beijing 100871, China}
\affiliation{Center for Quantitative Biology and Peking-Tsinghua Center for Life Sciences, AAIC, Peking University, Beijing 100871, China}
\author{Yuhai Tu}
\affiliation{IBM T. J. Watson Research Center, Yorktown Heights, New York 10598, USA}

\email{yuhai@us.ibm.com}
\maketitle

\section{Model description}
In this section, we describe the model studied in this paper in detail.
 As introduced in the main text, 
there are 3 types of molecules (3 species), denoted by $X_{1,2,3}$ respectively. The molecules can convert to different forms through 4 different reactions: 
\begin{gather}
X_1\autorightleftharpoons{$k_{12}$}{$k_{21}$}X_2, \label{eq:X12} \\
X_2\autorightleftharpoons{$k_{23}$}{$k_{32}$}X_3, \\
X_3\autorightleftharpoons{$k_{31}$}{$k_{13}$}X_1, \\
X_1+2X_2\autorightleftharpoons{$\tilde{k}_{12}$}{$\tilde{k}_{21}$}3X_2.\label{eq:X12coop}
\end{gather}
Basically, these molecules can convert to another form with a constant rate, and $X_1$ can transform to $X_2$ through an auto-catalytic reaction which introduces non-linearity to the system.

   
We consider this model a spatially-extended system with finite size (length) $L$. To describe the spatial distribution for each type of molecules, 
we divided the space into small ``boxes'' with size $\Delta x$ and count molecules $n_{1,2,3}(x)$ of the 3 types of molecules respectively in the box at each position $x$. The spatial profile of $n_i(x)$ or  the local concentration $u_i(x)\equiv \lim_{\Delta x\to0}n_i(x)/\Delta x$  can be used to describe the system's state. 

Naturally, the chemical reactions and the spatially stepping motion are stochastic processes which can be described by Markov processes. The forward and backward rates for above reactions are:
\begin{gather}
\left\{\begin{gathered}
rate[(n_1(x),n_2(x))\to(n_1(x)-1,n_2(x)+1)]=k_{12}n_1(x), \\
rate[(n_1(x),n_2(x))\to(n_1(x)+1,n_2(x)-1)]=k_{21}n_2(x),
\end{gathered}\right.\\
\left\{\begin{gathered}
rate[(n_2(x),n_3(x))\to(n_2(x)-1,n_3(x)+1)]=k_{23}n_2(x), \\
rate[(n_2(x),n_3(x))\to(n_2(x)+1,n_3(x)-1)]=k_{32}n_3(x),
\end{gathered}\right.\\
\left\{\begin{gathered}
rate[(n_3(x),n_1(x))\to(n_3(x)-1,n_1(x)+1)]=k_{31}n_3(x), \\
rate[(n_3(x),n_1(x))\to(n_3(x)+1,n_1(x)-1)]=k_{13}n_1(x),
\end{gathered}\right.\\
\left\{\begin{gathered}
rate[(n_1(x),n_2(x))\to(n_2(x)-1,n_3(x)+1)]=\tilde{k}_{12}^\prime n_1(x)n_2^2(x), \\
rate[(n_1(x),n_2(x))\to(n_2(x)+1,n_3(x)-1)]=\tilde{k}_{21}^\prime n_2^3(x),
\end{gathered}\right.
\end{gather}
where $\tilde{k}_{12}^\prime$ and $\tilde{k}_{21}^\prime$ are the rescaled rates: $\tilde{k}_{12}^\prime=\tilde{k}_{12}/(\Delta x)^2, \tilde{k}_{21}^\prime=\tilde{k}_{21}/(\Delta x)^2$.
The stepping rate between two boxes located at $x$ and $x+\Delta x$ (or $x-\Delta x$) respectively is considered to be proportional to the particle number difference, i.e., 
\begin{multline}
rate[(n_i(x),n_i(x\pm\Delta x))\to(n_i(x)-1,n_i(x\pm\Delta x)+1)]\\
=\left\{
\begin{aligned}
&\tilde{D}_i[n_i(x)-n_i(x\pm\Delta x)], n_i(x)>n_i(x\pm\Delta x); \\
&0, n_i(x)<n_i(x+\Delta x),
\end{aligned}\right.
\end{multline}
where $\tilde{D}_i$ is the stepping rate for $X_{i}$.
Within the framework of stochastic process, the dynamics of state variable $[(n_1(0), n_1(\Delta x),..., n_1(L)),(n_2(0), n_2(\Delta x),..., n_2(L)),(n_3(0), n_3(\Delta x),..., n_3(L))]$ can be directly simulated by Gillespie algorithm. 

In some cases, e.g., determining the Turing pattern bifurcation point, we care more about the mean-field dynamics than its noise.
In mean-field limit, where the noise of local concentration is ignored, the evolution of $u_i(x)$ is described by deterministic equations. For example, the dynamics of $u_1(x)$ can be described by
\begin{align}
\frac{\d u_1(x)}{\d t}=&-\tilde{k}_{12} u_1(x)u_2^{2}(x) +\tilde{k}_{21}u_2^{3}(x) + k_{21} u_2(x) -k_{12} u_1(x) - k_{13} u_1(x) + k_{31} u_3(x) \notag\\
&-\tilde{D}_1[n_1(x)-n_1(x-\Delta x)]+\tilde{D}_1[n_1(x+\Delta x)-n_1(x)]. \notag
\end{align}
In continuum limit where $\Delta x\to0$, by expanding $n_i(x\pm\Delta x)\approx n_i(x)\pm\partial_xn_i(x)\Delta x+\frac12\partial_x^2n_i(x)(\Delta x)^2$, these equations converge to PDEs:
	\begin{align}
	\frac{\partial u_1}{\partial t}&= -\tilde{k}_{12} u_1u_2^{2} +\tilde{k}_{21}u_2^{3} + k_{21} u_2 -k_{12} u_1 - k_{13} u_1 + k_{31} u_3 +D_{1}\frac{\partial^{2}u_1}{\partial x^{2}}, \\
	\frac{\partial u_2}{\partial t}&=\tilde{k}_{12} u_1u_2^{2} -\tilde{k}_{21}u_2^{3} - k_{21} u_2 +k_{12} u_1 + k_{32} u_3 - k_{23} u_2+D_{2}\frac{\partial^{2}u_1}{\partial x^{2}}, \\
	\frac{\partial u_3}{\partial t}&= -(k_{32}+k_{31}) u_3 +  k_{13} u_1 + k_{23}u_2 +D_{3}\frac{\partial^{2}u_3}{\partial x^{2}},
	\end{align}
where $D_i=\tilde{D}_i(\Delta x)^2$ is the diffusion constant. 

\section{Energy dissipation in spatially extended systems}
\subsection{The expression of energy dissipation}
In a spatially extended chemical reaction system, the free energy dissipation consists of two parts. One is due to local chemical reactions, and the other is due to spatial transport.

For local chemical reactions (reactions that happen within one spatial ``box''), the free energy dissipation rate is the same as spatially homogeneous system. For example, at position $x$, the dissipation rate within the ``box'' is
\begin{equation}
\dot{W}_\text{chem}(x)=\sum_i^{N_r}[J_i^+(x)-J_i^-(x)]\ln\left[\frac{J_i^+(x)}{J_i^-(x)}\right],
\end{equation} 
where $J_i^+(x)$ and $J_i^-(x)$ are the local forward flux and backward flux respectively of $i$-th reaction (in this model, there are $N_r=4$ reversible reactions) within each ``box''. For example, for the reaction  \eqref{eq:X12}, the forward flux is $J^+(x)=k_{12}n_1(x)$ and the backward flux is $J^-(x)=k_{21}n_2(x)$. The local dissipation rate {\it density} is thus
\begin{equation}
\dot{w}_\text{chem}(x)=\lim\limits_{\Delta x\to0}\dot{W}_\text{chem}(x)/\Delta x=\sum_i^{N_r}[j_i^+(x)-j_i^-(x)]\ln\left[\frac{j_i^+(x)}{j_i^-(x)}\right],
\end{equation} 
where $j_i^+(x)=\lim_{\Delta x\to0}J_i^+(x)/\Delta x$ and $j_i^-(x)=\lim_{\Delta x\to0}J_i^-(x)/\Delta x$ are the local forward flux and backward flux {\it density} (or the local concentration flux).

For dissipation from spatial transport, it can be calculated by regarding the molecule stepping between different boxes as jumping between different microscopic states. In this way, the forward and backward fluxes for specie $k$ between $x$ and $x+\Delta x$ are $J^+_{D,k}(x)=\tilde{D}_{k}n_k(x)$ and $J^-_{D,k}(x)=\tilde{D}_{k}n_k(x+\Delta x)$. In this way, by expanding $n_k(x+\Delta x)\approx n_k(x)\partial_xn_k(x)\times\Delta x$, the local dissipation rate is
\begin{equation}
\dot{W}_{\text{diff}}=\sum_{k=1}^3\frac{D_k(\frac{\partial n_k}{\partial x})^2}{n_k(x)},
\end{equation}
where $D_k=\tilde{D}_{k}(\Delta x)^2$. Thus, the local dissipation rate density is
\begin{equation}
\dot{w}_{\text{diff}}=\lim\limits_{\Delta x\to0}\dot{W}_{\text{diff}}/\Delta x= \sum_{k=1}^3 \frac{D_k(\frac{\partial u_k}{\partial x})^2}{u_k(x)}.
\end{equation}

Considering both two parts, the total dissipation rate of the system is
\begin{equation}\label{total}
\dot{W}=\int_0^L [\dot{w}_\text{diff}(x)+\dot{w}_\text{chem}(x)]\d x.
\end{equation}
\subsection{The break of detailed balance and the simplified expression of dissipation rate}
As discussed in main text, the ratio of the products of the reaction rates in the counter-clock wise and clockwise in these two respective chemical reaction cycles are:
 \begin{equation}
\Gamma=\frac{\tilde{k}_{21} k_{12}}{\tilde{k}_{12} k_{21}},\;\;\;\;
\Gamma^\prime= \frac{k_{13} k_{32} k_{21}}{k_{12} k_{23} k_{31}},
\end{equation}
which are both not equal to 1 and breaks detailed balance. The diffusion, however, is unbiased and thus does not involve any active processes or break detailed balance. Therefore, for all the microscopic cycles of microscopic states, only those consist chemical reaction subcycles would break detailed balance. Intuitively, only the parameters involving chemical reactions will explicitly appear in the expression of dissipation at steady state.

To formalize this discussion, we calculate \eqref{total} in details. Eq.(1) to Eq.(4) corresponds to the reaction subscript of $j_i^\pm(x)$ from $i=1,2,3,4$ respectively. Denote $j_i(x)=j_i^+(x)-j_i^-(x), j_{D,k}(x)=-D_k\partial_xu(x).$ At steady state, these local fluxes satisfy
\begin{align}
\frac{\partial j_{D,1}}{\partial x}&=-j_4-j_1+j_3, \\
\frac{\partial j_{D,2}}{\partial x}&=j_4+j_1-j_2, \\
\frac{\partial j_{D,3}}{\partial x}&=-j_3+j_2.
\end{align}
Writing out \eqref{total} explicitly and using the above equations, we have
\begin{align}
\dot{W}= &\int_0^L [\dot{w}_\text{diff}(x)+\dot{w}_\text{chem}(x)]\d x  \notag\\
= &\int_0^L \left[j_1\ln\left(\frac{k_{12}u_1}{k_{21}u_2}\right)
+j_2\ln\left(\frac{k_{23}u_2}{k_{32}u_3}\right)
+j_3\ln\left(\frac{k_{31}u_3}{k_{13}u_1}\right)
+j_4\ln\left(\frac{\tilde{k}_{12}u_1}{\tilde{k}_{21}u_2}\right)
+\sum_{k=1}^3 \frac{D_k(\frac{\partial u_k}{\partial x})^2}{u_k(x)}
\right]\d x \notag\\
=&\int_0^L \left[j_1\ln\left(\frac{k_{12}}{k_{21}}\right)
+j_2\ln\left(\frac{k_{23}}{k_{32}}\right)
+j_3\ln\left(\frac{k_{31}}{k_{13}}\right)
+j_4\ln\left(\frac{\tilde{k}_{12}}{\tilde{k}_{21}}\right)
\right. \notag \\
&\left.+(j_1-j_3+j_4)\ln u_1+(-j_1+j_2-j_4)\ln u_1+(-j_2+j_3)\ln u_3
+\sum_{k=1}^3 \frac{D_k(\frac{\partial u_k}{\partial x})^2}{u_k(x)}\right]\d x \notag\\
=&\int_0^L \left[j_1\ln\left(\frac{k_{12}}{k_{21}}\right)
+j_2\ln\left(\frac{k_{23}}{k_{32}}\right)
+j_3\ln\left(\frac{k_{31}}{k_{13}}\right)
+j_4\ln\left(\frac{\tilde{k}_{12}}{\tilde{k}_{21}}\right)
\right. \notag\\
&\left.
-\sum_{k=1}^3\frac{\partial j_{D,k}}{\partial x}\ln u_k
+\sum_{k=1}^3 \frac{j_{D,k}^2}{D_ku_k(x)}\right]\d x.
\end{align}
Since 
\[
\int_0^L\frac{\partial j_{D,k}}{\partial x}\ln u_k\d x=\int_0^Lj_{D,k}\d(\ln u_k)=\int_0^L \frac{j_{D,k}^2}{D_ku_k(x)}\d x,
\]
The simplified expression is
\begin{align}
\dot{W}&=
\int_0^L \left[j_1\ln\left(\frac{k_{12}}{k_{21}}\right)
+j_2\ln\left(\frac{k_{23}}{k_{32}}\right)
+j_3\ln\left(\frac{k_{31}}{k_{13}}\right)
+j_4\ln\left(\frac{\tilde{k}_{12}}{\tilde{k}_{21}}\right)\right]\d x \notag\\
&=J_{c1}\ln\Gamma^{-1}_1+J_{c_2}\ln\Gamma^{-1}_2,
\end{align}
where $J_{c1}=k_{21}N_2-k_{12}N_1$ and $J_{c2}=k_{23}N_2-k_{32}N_3$ ($N_i$ is the total molecule number of $X_i$ in the whole system) are the chemical cyclic flux in total in the respective cycles.
It appears that diffusion term doesn't appear in the simplified expression, which is in accordance with our previous intuitive understanding.

%

\section{Theoretical analysis of the dependence between positioning error and molecule number}
\subsection{How the onset depends on molecule number}
In this section, we introduce the theoretical analysis for the effect of varying total molecule number $N$. As mentioned in the main text, tuning molecule number would nonlinearly change the rate of the cooperative reaction \eqref{eq:X12coop}. To investigate this effect, we introduce the dimensionless (relative) local concentration $v_i(x)\equiv u_i(x)/u_{tot}$, where $u_{tot}=N/L$ is the averaged total concentration. The equations governing the mean-field dynamics of $v_i(x)$ are
\begin{align}\label{v1}
\frac{\partial v_1}{\partial t}&=
-\beta_{12}v_1v_2^2+\beta_{21}v_2^3
-k_{12}v_1+k_{21}v_2
-k_{13}v_1+k_{31}v_3
+D_1\frac{\partial^2 v_1}{\partial x^2}, \\
\label{v2}
\frac{\partial v_2}{\partial t}&=
\beta_{12}v_1v_2^2-\beta_{21}v_2^3
+k_{12}v_1-k_{21}v_2
-k_{23}v_2+k_{32}v_3
+D_2\frac{\partial^2 v_2}{\partial x^2}, \\
\label{v3}
\frac{\partial v_3}{\partial t}&=
k_{13}v_1-k_{31}v_3
+k_{23}v_2-k_{32}v_3
+D_3\frac{\partial^2 v_3}{\partial x^2}, 
\end{align}
where $\beta_{12}$ and $\beta_{21}$ are the normalized reaction rates of \eqref{eq:X12coop}: $\beta_{12}=\tilde{k}_{12}N^2/L^2, \beta_{21}=\tilde{k}_{21}N^2/L^2.$ These two rates are the function of $N$, and their ratio $\beta_{12}/\beta_{21}=\tilde{k}_{12}/\tilde{k}_{21}.$ As mentioned in the main text and previous section, there is a time scale separation, i.e.,  the rates related to $X_3$ are much smaller: $k_{13},k_{31},k_{23},k_{32}\ll k_{12},k_{21},\beta_{12},\beta_{21}.$ 

The homogeneous fixed-point solution $v_{1,2,3}(x)=v^*_{1,2,3}$ can be derived by setting the right-hand side (RHS) of the above equations to be zero, which satisfies
\begin{gather}\label{N:eq1}
-\beta_{12}v_1^*v_2^{*2}+\beta_{21}v_2^{*3}
-k_{12}v_1^*+k_{21}v_2^*
-k_{13}v_1^*+k_{31}v_3^*
=0, \\
\label{N:eq2}
\beta_{12}v_1^*v_2^{*2}-\beta_{21}v_2^{*3}
+k_{12}v_1^*-k_{21}v_2^*
-k_{23}v_2^*+k_{32}v_3^*
=0, \\
\label{N:eq3}
k_{13}v_1^*-k_{31}v_3^*
+k_{23}v_2^*-k_{32}v_3^*
=0.
\end{gather}
From \eqref{N:eq3} we have
\[
v_3^*=\frac{k_{13}v_1^*+k_{23}v_2^*}{k_{31}+k_{32}},
\]
so we can eliminate $v_3^*$ and derive the constraints between $v_1$ and $v_2$:
\begin{equation}\label{v12}
-\beta_{12}v_1^*v_2^{*2}+\beta_{21}v_2^{*3}
-k_{12}v_1^*+k_{21}v_2^*
=0,
\end{equation}
where we dropped the $u_3^*$ terms because they are small.
In addition, the conservation condition $v_1^*+v_2^*+v_3^*=1$ gives
\begin{equation}
R_1v_1^*+R_2v_2^*=1,\label{conserve}
\end{equation}
where
\[
R_1=\frac{k_{31}+k_{32}+k_{13}}{k_{31}+k_{32}}>1,R_2=\frac{k_{31}+k_{32}+k_{23}}{k_{31}+k_{32}}>1,
\]
are the  ratios of the rates related to $X_3$. Eqs. (\ref{v12}) and (\ref{conserve}) determine $v^*_{1,2}$ together.

Next, we apply linear analysis near this homogeneous fixed point. Due to the separate time scale, $v_1$ and $v_2$ changes much faster than $v_3$ near the fixed point. Hence we can 
perturb $v_1$ and $v_2$ in \eqref{N:eq1} and \eqref{N:eq2} near the homogeneous fixed point: $v_j(x;q)=v_j^*+\delta v_j\times\e^{\i qx}, j=1,2,$ while keeping $v_3=v_3^*$ constant during the perturbation.
Substituting into Eqs.(\ref{v1}-\ref{v3}), we can calculate the criteria to have a positive eigenvalue for pattern with wave number $q$:
\begin{equation}
-q^2[D_1(2\beta_{12}v_1^*v_2^*-3\beta_{21}v_2^{*2}-k_{21})+D_2(-\beta_{12}v_2^{*2}-k_{12} )]+q^4D_1D_2<0,
\end{equation}
from which we can determine the criteria for the onset of Turing pattern in infinite system,
\begin{equation}
C\equiv(2\beta_{12}v_1^*v_2^*-3\beta_{21}v_2^{*2}-k_{21})+d^{-1}(-\beta_{12}v_2^{*2}-k_{12} )>0, \label{qc}
\end{equation}
with critical wave number $q_c=\sqrt{C/(2D_2)}$. Equivalently, this leads to
\begin{equation}\label{onset}
\beta_{21}<\beta_0\equiv\frac1{3v_2^{*2}}[(2\beta_{12}v_1^*v_2^*-k_{21})-d^{-1}(\beta_{12}v_2^{*2}+k_{12} )],
\end{equation}
with $\beta_0$ the critical rate of $\beta_{21}$. 
This equation indicates that $\beta_{21}$ should be smaller than a critical value $\beta_0$  for Turing pattern to arise.
Combining with \eqref{v12} and considering $d\gg1$, $\beta_0$ can also be expressed by
\begin{equation}\label{onset2}
\beta_{0}=-2\frac{k_{12}}{R_1}\frac{1}{v_2^{*3}}+\left(2k_{12}\frac{R_2}{R_1}+k_{21}\right)\frac{1}{v_2^{*2}},
\end{equation}
For given $\tilde{k}_{12}$ and $N$, the solution of $v_1^*, v_2^*$ and $\beta_0$ can be determined from Eqs.(\ref{v12}), (\ref{conserve}) and (\ref{onset2}).
In this way, we can calculate the critical $\Gamma$ at the onset in infinite system $\Gamma_0=\tilde{k}_{0}k_{12}/(\tilde{k}_{12}k_{21})=\beta_0k_{12}/(\beta_{12}k_{21}).$ Only for those $\Gamma<\Gamma_0$, would the Turing pattern be possible.

When $N$ varies, both $v_1^*$ and $v_2^*$ will be changed, and so will $\beta_0$.
From \eqref{onset}, $\beta_0$ implicitly depends on $N$ through $v_2^*$ and, by calculating the derivative of the function $\beta_0(v_2^*)$, it can be checked that, for typical kinetic rates $\tilde{k}_{12}\gg\tilde{k}_{21}$ and $k_{21}\gg k_{12}$, the $\beta_0$ dependence on $v_{2}^*$ is non-monotonic and there is a maximum $\beta_0$ at
 \begin{equation}
 v_2^*=\frac32\left(R_2+\frac12\frac{k_{21}}{k_{12}}R_1\right)^{-1}\equiv v_{2,m}^*,
  \end{equation}
and the corresponding maximum $\beta_{0}$ is
\begin{equation}\label{betaCmax}
\beta_{0,max}=\frac{4}{27}
\frac{(\frac{2k_{12}R_2}{R_1}+k_{21})^3}
{(\frac{2k_{12}}{R_1})^2},
\end{equation}
which only depends on kinetic rates of linear reactions.

Hence, $\Gamma_0$ would vary with $N$ in a non-monotonic way, which would effectively tune the control parameter $\epsilon$ and lead to a non-monotonic dependence of the positioning error on $N$.

\subsection{The analysis of $N_{max}$ and the relation between error and molecule number in the case of large $N$}
When $N$ is large, we assume $\beta_{12}$ is much larger than all the other rates in the system and consequently $v_2\gg v_1$. In this case, we can derive the approximate relation between positioning error $\sigma$ and $N$, and find out how the key parameters (e.g. irreversible parameter $\Gamma$ and diffusion rate ratio $d$) determines $N_{max}$.

\eqref{conserve} gives 
\[
v_2^*=\frac1{R_2}(1-R_1v_1^*).
\]
Since $v_1^*\ll v_2^*$, substituting this expression into \eqref{v12} and only keeping the leading terms, we have
\begin{equation}\label{v1*}
v_1^*=\left(R_2k_{21}+\frac{\beta_{21}}{R_2}\right)\frac1{\beta_{12}},
\end{equation}
where we have considered $\beta_{12}$ much larger than other rates. With these approximate expression of $v_{1,2}^*$, $\beta_{0}$ can be expressed explicitly:
\begin{equation}
\beta_{0}=k_{21}R_2^2-\frac{\beta_{12}}{d}.
\end{equation}
Thus, the critical $\Gamma$ in infinite system is
\begin{equation}
\Gamma_0=\frac{\beta_0k_{12}}{\beta_{12}k_{21}}
=\frac{k_{12}R_2^2}{\beta_{12}}-\frac{k_{12}}{k_{21}d}
=\frac{k_{12}R_2^2L^2}{\tilde{k}_{12}}\frac1{N^2}-\frac{k_{12}}{k_{21}d}.
\end{equation}
Since $\Gamma_c\propto\Gamma_0$, there should be a similar relation between $\Gamma_c$ and $N$:
\begin{equation}
\Gamma_c=\frac{\tilde\alpha}{N^{2}}-\frac{\tilde\alpha^\prime}d,
\end{equation}
with $\tilde\alpha_{1,2}$ two coefficients determined by the kinetic rates. 

From this relation, we can also determine how $N_{max}$ depends on $\Gamma$ and $d$. When $N=N_{max}$, system is exactly at critical point, i.e., $\epsilon=0$, which gives
\begin{equation}\label{Nmax}
\Gamma_c=\frac{\tilde\alpha}{N_{max}^{2}}-\frac{\tilde\alpha^\prime}d=\Gamma\Rightarrow N_{max}
=\left(\frac{\tilde{\alpha}}{\Gamma+\tilde\alpha^\prime d^{-1}}\right)^{\frac12}.
\end{equation}
\eqref{Nmax} shows how $N_{max}$ depends on $\Gamma$ and $d$. In usual cases where $\Gamma$ is finite and $d\gg1$, $N_{max}\propto\Gamma^{-1/2}=\e^{W/2}$, where $W\equiv-\ln\Gamma$ is the energy of the pattern. It indicates that exponentially increasing pattern's energy cost could also increase $N_{max}$ and thus broaden the available range of $N$. This relation is tested in the main text. On the other hand, there is an overall upper limit of $N_{max}$ determined by diffusion rate ratio: when energy cost is so large that $\Gamma\ll\tilde\alpha^\prime d^{-1}$, $N_{max}$ is dominated by $N_{max}\propto d^{1/2}$.

%


\section{The energy dissipation for overcoming diffusion is a small fraction of the total energy dissipation}
In Fig.2 of the main text, we show that the proportion of energy dissipated to overcome diffusion is quite low, i.e., $r_{diff}$, which holds true even deep in the Turing pattern regime, e.g., for large $d$. Here, we provide an estimation of $r_{diff}$ to show the generality of this result.
%
     
    Since $X_2$ shows the most significant spatial variation, its diffusion energy dissipation is the main contribution to the overall diffusion energy dissipation, i.e, $\dot{w}_{diff}\approx{D_2(\frac{\partial u_2}{\partial x})^2}/{u_2(x)}.$ The derivative $\frac{\partial u_2}{\partial x}$ can be estimated by $(2\Delta u_2)/(\lambda/2)$, where $\lambda$ is the wavelength of the pattern and $\Delta u_2\equiv(u_{2,max}-u_{2,min})/2$ is the amplitude of the pattern. By considering \eqref{qc}, diffusion energy dissipation rate density can be estimated by 
\[
\dot{w}_{diff}\approx\frac{16D_2(\Delta u_2)^2q_c^2}{4\pi^2u_2}
\approx\frac{2(2\beta_{12}v_1^*v_2^*-k_{21})(\Delta u_2)^2}{\pi^2u_2}.
\]
 where we consider $\beta_{21}$ and $d^{-1}$ to be small and drop these terms. From \eqref{v12} (also considering $\beta_{21}$ and $k_{12}v_1^*$ to be small), $\beta_{12}v_1^*v_2^*\approx k_{21}$, thus eventually
 \begin{equation}
\dot{w}_{diff}
\approx\frac{2k_{21}(\Delta u_2)^2}{\pi^2u_2}.
\end{equation}
    
    For chemical reaction dissipation rate, since the cyclic flux $J_{1c}\gg J_{2c}$, $\dot{w}_{chem}$ is approximated by $\dot{w}_{chem}\approx(k_{21}u_2-k_{12}u_1)\ln\Gamma^{-1}\approx k_{21}u_2\ln\Gamma^{-1}$. Hence the overall ratio between diffusion dissipation and chemical reaction dissipation is
    \begin{equation}
    \frac{\dot{W}_{diff}}{\dot{W}_{chem}}
    =\frac{\int\dot{w}_{diff}\d x}{\int\dot{w}_{chem}\d x}
    \approx
    \frac{2}{\pi^2\ln\Gamma^{-1}}\left(\frac{\Delta u_2}{\langle u_2 \rangle}\right)^2,
    \end{equation}
where $\langle u_2 \rangle\equiv L^{-1}\int_0^Lu_2(x)\d x$ is the averaged concentration over space. For patterns far from onset, $\ln(\Gamma^{-1})>5$ and the prefactor $2/[\pi^2\ln(\Gamma^{-1})]< 0.05$ is quite small. Even for patterns far from the onset where $\Delta u_2$ is of the same order of $\langle u_2 \rangle$, $\dot{W}_{diff}$ is still much smaller than $\dot{W}_{chem}$, which should hold in general in this system.

\section{The role of $X_3$}
In the most part of the paper, we considered that the reactions related to $X_3$ are much slower, and these reactions are ignored in the theoretical analysis. However, it doesn't indicate that $X_3$ is useless in the system. Actually, although it doesn't involve in the pattern formation process, it contributes a lot to the long-term stability of the Turing pattern.

 \begin{figure}
       	\centering
       	\includegraphics[width=1\linewidth]{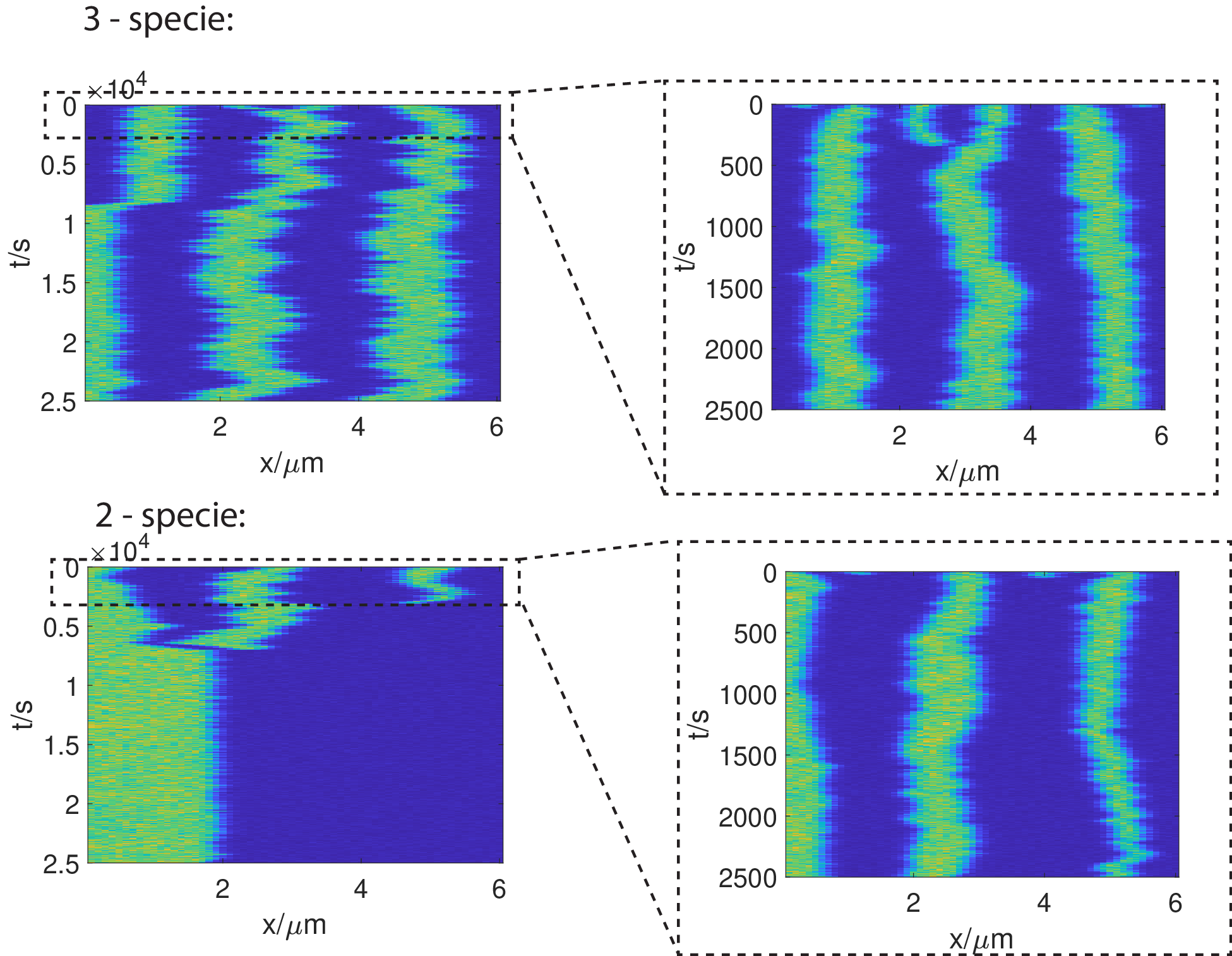}
       	\caption{ Kymographs of $X_2$ for 2-specie model and 3-specie model with same parameters. Although their short-term dynamics are similar, the long-term dynamics are quite different.}
\label{X3}
       \end{figure}

Fig.\ref{X3}(A) and (B) show the kymographs with or without $X_3$ reactions respectively. It turns out that their short-term dynamics ($t\ll T_{short}\approx2500$) are similar, but their long-term dynamics are quite different. In 3-specie case, the pattern always has the same wave number (in the Figure there are always three stripes), with each stripe fluctuating around its mean position. In 2-specie case, although at first the stripes also fluctuating around a fixed position, the wavenumber of the pattern will change over a long time scale. This long-term instability of pattern with specific wavenumber will lead to the ambiguity of mean peak position of stripes, and is harmful for spatial positioning in biological systems. 

In addition, it should be noted that the result of this instability appears in a  much longer time scale than the time scale of pattern formation and the correlation time of th stripe peak fluctuation. 
Hence, intuitively the mechanism to fight against this instability should also has a much slower rate than pattern formation and peak fluctution.
This might explain that the rates of $X_3$ reactions can be much smaller than other reactions: it works in a much different time scale.

In our theoretical analysis of pattern, we mainly considered the realistic case where the pattern has a fixed wavenumber. This requires the pattern to be pinned at the right place for a long time, which is fulfilled by $X_3$-related dynamics.  
Therefore, the characterization of long-term positioning error (i.e., assuming a fixed wavenumber and relating positioning error with the phase fluctiation) has taken the role of $X_3$ into consideration, and $X_3$ is not totally neglected from the theoretical analysis of the model.  

\section{Our model of Muk system}
In the last section in main text, we briefly introduced a realistic biochemical system, namely MukBEF system, to illustrate how our theory applies and can be tested by experiments. Here we introduce the details of the model we use. 

According to the background introduced in main text, we use a biochemical reaction-diffusion network to model the working mechanism of Muk system, as is shown in . $X_2$ and $X_1$ are the closed dimer of dimer that captures the DNA or not respectively, while $X_3$ represents open MukBEF dimer. $M_1$ and $M_2$ are two intermediate states right after each corresponding ATP hydrolysis step. The red lines in the figure represents those reactions hydrolyzing ATP, and the reverse rates are regulated by ADP concentration. $X_1, X_3$ and $M_2$ are assumed to have the same faster diffusion constant $D_1$, whereas $M_1$ and $X_2$ are assumed to have the same slower diffusion constant $D_2$. The reactions between different molecules are:
\begin{gather}
X_1+2M_1\autorightleftharpoons{${k}_1^+$}{${k}_1^-[ADP]$}3M_1, \\
M_1\autorightleftharpoons{${k}_2^+[ATP]$}{${k}_2^-$}X_2, \\
X_2\autorightleftharpoons{${k}_3^+$}{${k}_3^-[ADP]$}M_2, \\
M_2\autorightleftharpoons{${k}_4^+[ATP]$}{${k}_4^-$}X_1, \\
X_2\autorightleftharpoons{$k_{23}$}{$k_{32}$}X_3, \\
X_3\autorightleftharpoons{$k_{31}$}{$k_{13}$}X_1.
\end{gather}


With this model, as discussed in the main text, many predictions in realistic systems can be made, e.g., how positioning behavior will change when $[ATP]/[ADP]$ ratio is changed. The simulation is constructed in the same way introduced in Section I. The parameters are:  
$k_{1}^{+}=1.67\times 10^{-5} \text{s}^{-1}\mu\text{m}^{2}$, $k_{1}^{-}[ADP]_0=3.34\times 10^{-6} \text{s}^{-1}\mu\text{m}^{2}$,
 $k_{2}^{+} [ATP]_0=20 \text{s}^{-1} $, $k_{2}^{-}=10 \text{s}^{-1}$, 
 $k_{3}^{+}=3.6 \text{s}^{-1}$, $k_{3}^{-} [ADP]_0=2.5 \text{s}^{-1}$,
  $k_{4}^{+}[ATP]_0=20 \text{s}^{-1} $, $k_{4}^{-}=4 \text{s}^{-1}$ , 
   $D_1=0.3 \mu\text{m}^{2}\text{s}^{-1}$, $D_2=0.012 \mu\text{m}^{2}\text{s}^{-1}$, where $[ATP]_0$ and $[ADP]_0$ are the ATP or ADP concentration in standard condition respectively. 
We tuned the reverse rates of the red reactions in Fig.5 in main text to simulate the change of ADP concentration and, consequently, the change of $[ATP]/[ADP]$ ratio.

\bibliographystyle{unsrt}
\bibliography{ref.bib}